\documentclass[3p,times,nopreprintline]{elsarticle}

\usepackage{microtype}
\usepackage{hyperref}
\biboptions{sort&compress}
\usepackage[dvipsnames]{xcolor}
\usepackage[inline]{enumitem}
\usepackage{amsmath}
\usepackage{amssymb}
\usepackage[T1]{fontenc}
\usepackage{newtxtext,newtxmath}
\usepackage{cleveref}
\usepackage[font=normalsize,labelfont=bf]{caption}
\usepackage{subcaption}
\usepackage{xspace}

\hypersetup{
    colorlinks,
    linkcolor={red!50!black},
    citecolor={blue!50!black},
    urlcolor={blue!80!black}
}

\setlist{label=(\textit{\roman*})}

\usepackage{tikz}
\usetikzlibrary{shapes.geometric, arrows,positioning,fit,scopes,backgrounds}
\tikzstyle{box} = [rectangle, text width=4cm, minimum height=1cm, text centered, draw=black, fill=red!30, thick, inner sep=5pt]
\tikzstyle{arrow} = [very thick,->,>=stealth]
\tikzstyle{line} = [very thick]
\tikzstyle{codebox} = [rectangle, text width=3.cm, minimum height=1cm, text centered, draw=black, fill=green!30, thick, inner sep=5pt]
\tikzstyle{gf} = [draw=black, thick, fill=blue!75!black!25, inner sep=8pt]
\tikzstyle{expbox} = [rectangle, text width=4cm, minimum height=1cm, text centered, draw=black, fill=orange!30, thick, inner sep=5pt]
\tikzstyle{sumpbox} = [rectangle, text width=3.1cm, minimum height=1cm, text centered, draw=black, fill=purple!30, thick, inner sep=5pt]

\newcommand{\code}[1]{\textsf{#1}}
\newcommand{\tev}{\,\text{TeV}}
\newcommand{\gev}{\,\text{GeV}}
\newcommand{\like}{\mathcal{L}}
\newcommand{\prob}{P}
\newcommand{\given}{\,|\,}
\newcommand{\data}{\textbf{D}}
\newcommand{\params}{\textbf{x}}
\newcommand{\see}[1]{(see e.g.~refs.~\cite{#1})}
\newcommand{\seeone}[1]{(see e.g.~\refcite{#1})}
\newcommand{\refcite}[1]{ref.~\cite{#1}}
\newcommand{\refscite}[1]{refs.~\cite{#1}}
\newcommand{\Refcite}[1]{Ref.~\cite{#1}}
\newcommand{\Refscite}[1]{Refs.~\cite{#1}}
\newcommand{\gambit}{\code{GAMBIT}}
\DeclareMathOperator{\sign}{sign}
\newcommand{\CP}{\ensuremath{\mathcal{CP}}\xspace}

\begin{document}

\begin{frontmatter}

	\title{Global fits and the search for new physics: past, present and future\\[3mm]{\normalsize\bf The \gambit{} Collaboration}\\[-5mm]}

	\author[a,b]{Peter Athron}
	\ead{peter.athron@njnu.edu.cn}
	\author[c]{Csaba Bal\'azs}
	\author[s]{Jon Butterworth}
	\author[h]{Christopher Chang}
	\author[d]{Andrew Fowlie}
	\ead{andrew.fowlie@xjtlu.edu.cn}
	\author[e]{Tom\'as Gonzalo}
	\author[f,g]{Adil Jueid}
	\ead{adil.hep@gmail.com}
	\author[h]{Anders Kvellestad}
	\author[i,j]{Michele Lucente}
	\author[k,l,m]{Farvah Mahmoudi}
    \author[z]{Gregory D. Martinez}
	\author[h]{Are Raklev}
	\author[n]{Roberto Ruiz de Austri}
	\author[o]{Cristian Sierra}
	\author[p]{Wei Su}
	\author[q]{Aaron~C.~Vincent}
	\author[r]{Martin White}
	\author[a,b]{Lei Wu}

	\address[a]{Department of Physics and Institute of Theoretical Physics, Nanjing Normal University, Nanjing, 210023, China}
	\address[b]{Nanjing Key Laboratory of Particle Physics and Astrophysics, Nanjing, 210023, China}
	\address[c]{School of Physics and Astronomy, Monash University, Melbourne, VIC 3800, Australia}
	\address[s]{Department of Physics and Astronomy, University College London, London WC1E 6BT, UK}
	\address[h]{Department of Physics, University of Oslo, N-0316 Oslo, Norway}
	\address[d]{X-HEP Laboratory, Department of Physics, School of Mathematics and Physics, Xi'an Jiaotong-Liverpool University, \\ 111 Ren'ai Road, Suzhou Dushu Lake,
		Science and Education Innovation District, Suzhou Industrial Park, Suzhou 215123, China}
	\address[e]{Institut f\"ur Theoretische Teilchenphysik, Karlsruher Institut f\"ur Technologie (KIT), D-76128 Karlsruhe, Germany}
	\address[f]{Particle Theory and Cosmology Group, Center for Theoretical Physics of the Universe, \\ Institute for Basic Science (IBS), 34126 Daejeon, Republic of Korea}
	\address[g]{Cosmology, Gravitation and Astroparticle Physics Group, Center for Theoretical Physics of the Universe, \\ Institute for Basic Science (IBS), 34126 Daejeon, Republic of Korea}
	\address[i]{Dipartimento di Fisica e Astronomia, Universit\`a di Bologna, via Irnerio 46, 40126 Bologna, Italy}
	\address[j]{INFN, Sezione di Bologna, viale Berti Pichat 6/2, 40127, Bologna, Italy}
	\address[k]{Universit\'e Claude Bernard Lyon 1, CNRS/IN2P3, \\ Institut de Physique des 2 Infinis de Lyon, UMR 5822, F-69622, Villeurbanne, France}
	\address[l]{Theoretical Physics Department, CERN, CH-1211 Geneva 23, Switzerland}
	\address[m]{Institut Universitaire de France (IUF), 75005 Paris, France}
    \address[z]{Department of Physics and Astronomy, UCLA, Los Angeles, CA 90095-1547, USA}
	\address[n]{Instituto de F\'{\i}sica Corpuscular, CSIC-Universitat de Val\`encia, E-46980 Paterna, Valencia, Spain}
	\address[o]{Tsung-Dao Lee Institute \& School of Physics and Astronomy, Shanghai Jiao Tong University, Shanghai 200240, China}
	\address[p]{School of Science, Shenzhen Campus of Sun Yat-sen University, \\ No.\ 66, Gongchang Road, Guangming District, Shenzhen, Guangdong 518107, China}
	\address[q]{Department of Physics, Engineering Physics and Astronomy,
		Queen's University, Kingston ON K7L 3N6, Canada}
	\address[r]{ARC Centre of Excellence for Dark Matter Particle Physics \& CSSM, \\
		Department of Physics, University of Adelaide, Adelaide, SA 5005, Australia}

	\begin{abstract}
		In this work, we review the history and current role of global fits in the search for physics beyond the Standard Model~(BSM), including precision tests of the Standard Model (SM).
		Although BSM global fits were initially focused on minimal supersymmetric models, we describe how  fits have evolved in response to new data from the Large Hadron Collider (LHC) and elsewhere, expanding to encompass a broad spectrum of BSM scenarios including non-minimal supersymmetry, axion-like particles, extended Higgs sectors, dark matter models, and effective field theories such as SMEFT.
		We discuss how the role of global fits has shifted from forecasting possible signals of new physics at the LHC to understanding the impact of null results from LHC run-I and II and the discovery of the Higgs boson, and how interest has shifted from global fits for parameter estimation to comprehensive model comparison.
		We close by discussing potential trends and future applications, emphasizing the potential for machine learning and artificial intelligence to enhance the efficiency of sampling algorithms and comparison between theory and experiment, as well as collaboration and software development.
	\end{abstract}

\end{frontmatter}

\clearpage
\tableofcontents

\section{Introduction}\label{sec:intro}

Global fits play an important role in assessing scientific models and their predictions by combining all relevant data from many different experiments, intelligently sampling over the relevant parameters
and rigorously applying statistical methods to fit them to the combined data.  This process enables significantly more
science to be extracted
than from naive approaches to combining experimental data, and as a result it maximizes the impact of scientific experiments~\cite{AbdusSalam:2020rdj}. In particle physics and the search for new physics this is particularly important due to the high cost of building and operating experiments, as well as the large set of models and parameter spaces characterizing new physics possibilities.  In this review, we discuss the important role of global fits in the search for new physics, and how that role continues to evolve during the Large Hadron Collider (LHC) era. We put particular emphasis on the impact of the Higgs boson discovery~\cite{ATLAS:2012yve,CMS:2012qbp}, the absence of other new physics signatures at the LHC, and the emergence of machine learning~(ML) in particle physics~\cite{Feickert:2021ajf}.

Despite the tremendous experimental successes of the Standard Model (SM), it does not address decades-old puzzles such as the nature of dark matter (DM), the origin of neutrino masses, the stability of the weak scale, or the origin of the matter-antimatter asymmetry in the Universe.  These puzzles motivate proposing and searching for physics beyond the SM, such as new fundamental particles and symmetries.  The search strategies include direct searches at high-energy collider experiments such as the LHC~\cite{Morrissey:2009tf};
direct searches for DM in underground detectors~\cite{Billard:2021uyg}; indirect searches for DM at neutrino and gamma-ray telescopes~\cite{Gaskins:2016cha};
measurements of electric and magnetic dipole moments~\cite{Athron:2025ets};
axion searches such as haloscopes, helioscopes~\cite{Irastorza:2018dyq}
and light shining through a wall~\cite{Redondo:2010dp};
gravitational wave observatories~\cite{Athron:2023xlk};
and $B$-factories such as Belle and BaBar~\cite{BaBar:2014omp}.
In light of data from so many searches,
global fits are an indispensable tool to combine the data, shape our understanding of new physics beyond the SM, and formulate strategies for future discoveries.

\begin{figure}
	\centering
	\begin{tikzpicture}[node distance=0.8cm]
		\node (th) [box] {New physics theory, e.g.~Lagrangian, $\mathcal{L}$};
		\node (symbolic) [codebox, below=of th] {\code{GUM}~\cite{Bloor:2021gtp}, \code{SARAH}~\cite{Staub:2010jh,Staub:2012pb,Staub:2013tta,Goodsell:2015ira}, \code{FeynRules}~\cite{Christensen:2008py,Alloul:2013bka}, \code{MARTY}~\cite{Uhlrich:2020ltd} etc};
		\node (model) [box, below=of symbolic] {Computer representation of model};
		\node (codes) [codebox, below=of model] {\code{SOFTSUSY}~\cite{Allanach:2001kg}, \code{micrOMEGAs}~\cite{Belanger:2001fz,Belanger:2006is,Alguero:2023zol}, \code{Pythia}~\cite{Sjostrand:2006za} etc};
		\node (obs) [box, below=of codes] {Model predictions};
		\node (statmodel) [box, below=of obs] {Statistical model
			\& likelihood $\mathcal{L}$};
		\node (algo) [codebox, below=of statmodel] {Sampling algorithm --- \code{Diver}~\cite{Martinez:2017lzg}, \code{PolyChord}~\cite{Handley:2015fda} etc};
		\node (res) [box, below=of algo] {Statistical inferences};
		\node (conc) [box, below=of res] {Scientific conclusions};
		\node (exp) [expbox, right=of th] {Data, likelihoods, statistical models, etc.\ from all relevant experimental searches};
		\node (statchoices) [sumpbox, left=of res] {Choice of statistical framework};
		\node (sm) [sumpbox, left=of statmodel] {Assumptions about backgrounds};
		\draw [line] (th) -- (symbolic);
		\draw [arrow] (symbolic) -- (model);
		\draw [line] (model) -- (codes);
		\draw [arrow] (codes) -- (obs);
		\draw [arrow] (obs) -- (statmodel);
		\draw [arrow] (exp) |- (statmodel);
		\draw [arrow] (sm) -- (statmodel);
		\draw [line] (statmodel) -- (algo);
		\draw [arrow] (algo) -- (res);
		\draw [arrow] (statchoices) -- (res);
		\draw [arrow] (res) -- (conc);
		\begin{scope}[on background layer]
			\node [gf, fit=(symbolic) (statchoices) (statmodel)] {};
		\end{scope}
	\end{tikzpicture}
	\medskip
	\caption{Sketch of the elements of a global fit in particle physics --- through the fit (purple box), we combine a new physics theory and experimental searches, starting from the Lagrangian and going all the way to scientific conclusions. Red boxes show the key elements and green boxes show examples of computer programs that take them as inputs and deliver them as outputs.}
	\label{fig:sketch}
\end{figure}
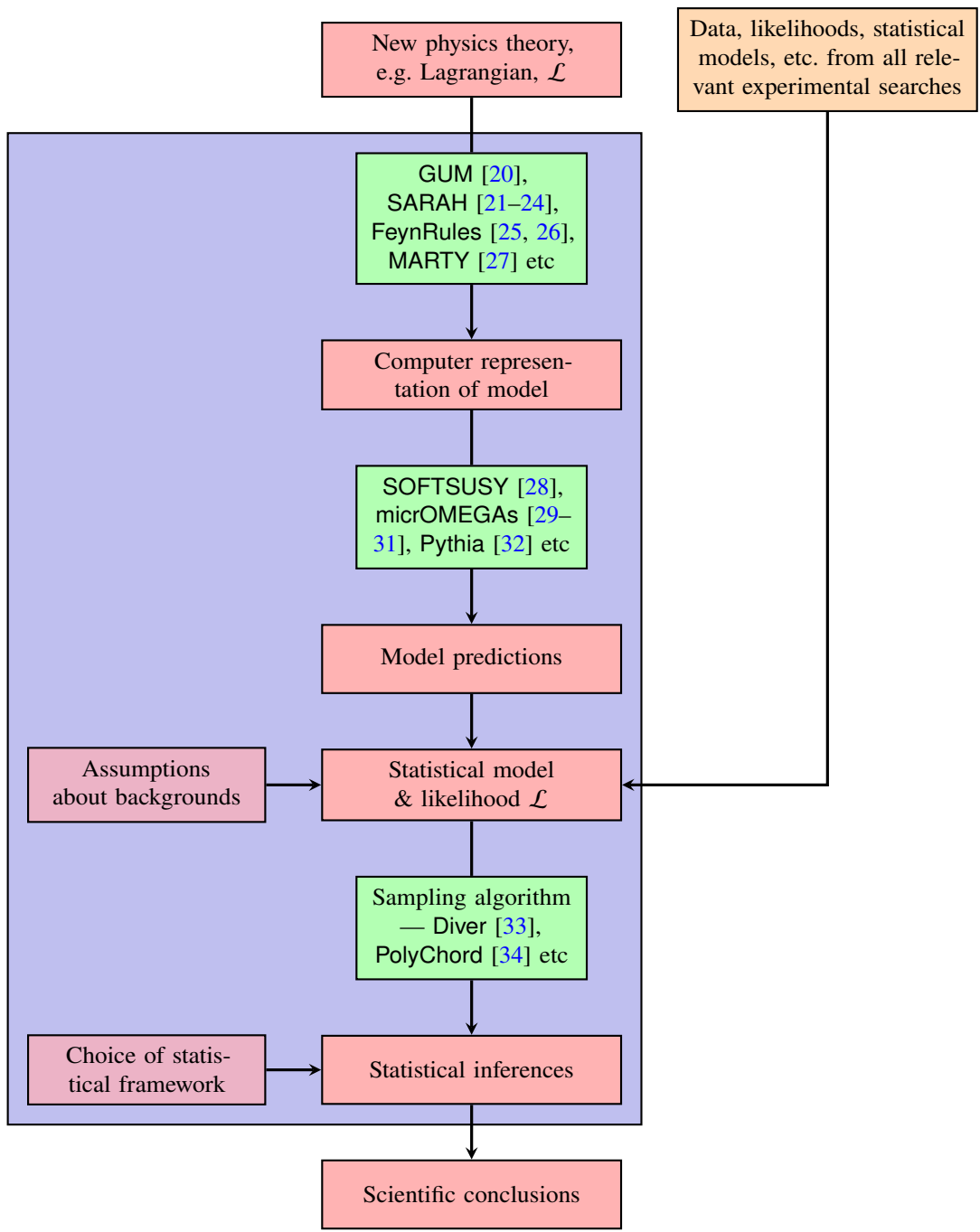

In a global fit one fits all a model's parameters, including any nuisance parameters such as the top mass,  to all relevant available experimental data, with two distinct goals. First, \emph{parameter estimation}, where one determines which parameter sets are best able to explain the observed data within a given theoretical model.\footnote{These global fits generally do check at least that the best-fit point is consistent with the observables in the fit, or comment on any observables that are in tension.}
Second, \emph{model testing} and \emph{model comparison}, where one tests or compares models' abilities to fit the observed experimental data. These two goals can be formulated in two different statistical perspectives:
\begin{enumerate*}
	\item frequentist statistics, where one is concerned about controlling the long-run frequency of errors \seeone{Eadie:100342}, and
	\item Bayesian statistics, where one attempts to quantify the plausibility of parameter choices and models using probability theory \seeone{2005blda.book.....G}.
\end{enumerate*}
The \emph{likelihood function}~\cite{Cousins:2020ntk} --- the probability or probability density of data given a particular model and parameters
\begin{equation}
	\like(\params) \equiv \prob(\data \given M, \params)
\end{equation}
is a central component in both approaches (though see likelihood free methods~\cite{Brehmer:2020cvb}). These likelihoods reflect our modelling of the data-generating process. Thus global fitting usually involves careful (re)construction of the likelihood, though in recent years there has been a push towards publicly available likelihoods and statistical models~\cite{LHCReinterpretationForum:2020xtr,Cranmer:2021urp,LHCReinterpretationForum:2025zgq}. Having constructed a likelihood, one proceeds to statistical inference, which can be computationally demanding for models with many parameters. As we shall discuss, the computational methods available have evolved over the last twenty years, particularly in the last few years with the advent of ML methods. The outline of a global fit in the context of particle physics is sketched in \cref{fig:sketch}.

Beginning in the early 1990s, global fits were used to perform precision tests of the SM using Tevatron, SLC and LEP data, with poor fits potentially providing evidence for new physics and clues about how to find it~\cite{Altarelli:1997et,Altarelli:2004fq,Erler:2019hds}.
Global fits to forecast the masses and couplings of new particles within reach of the LHC and Tevatron were carried out in the early 2000s in both minimal~\cite{Baer:2003yh,Ellis:2003si,Profumo:2004at,Baltz:2004aw,Allanach:2005kz,RuizdeAustri:2006iwb,Roszkowski:2006mi,Roszkowski:2007fd,Allanach:2007qk,Cabrera:2009dm,Buchmueller:2007zk,Buchmueller:2008qe,Bechtle:2010igv} and next-to-minimal~\cite{Balazs:2009su,Lopez-Fogliani:2009qdp,Kowalska:2012gs,Fowlie:2014faa} supersymmetric models.
After the LHC began running later global fits showed the impact that the LHC was having on minimal supersymmetry, and, if minimal supersymmetry is taken as a benchmark model, BSM physics in general.  Global fits have since broadened into other models, including increased interest in non-minimal variants of supersymmetry~\cite{Cao:2013gba,Cheung:2014lqa,King:2014xwa,Abdughani:2021pdc}, models of axions and axion-like particles~\cite{Hoof:2018ieb,Athron:2020maw,Balazs:2022tjl}, two Higgs doublet models (2HDMs)~\cite{Chowdhury:2015yja,Karan:2023kyj,Athron:2021auq,Beniwal:2022kyv,Athron:2024rir}, Higgs portal DM~\cite{GAMBIT:2017gge,Athron:2018ipf,GAMBIT:2018eea} and various effective~\cite{GAMBIT:2021rlp} or simplified models of DM~\cite{Chang:2022jgo,Chang:2023cki,Bagnaschi:2019djj}. The use of global fits for model comparisons also increased compared to the early studies that were exclusively concerned with parameter estimation. Lastly, recent tough constraints on light new physics at the LHC have motivated more studies on effective field theories, such as Standard Model Effective Field Theory (SMEFT)~\cite{Ellis:2018gqa,deBlas:2022ofj}, where new physics effects are captured in higher-dimensional operators.

With the LHC run III expected to close by July 2026, and high-luminosity running anticipated to commence near the end of the decade, this review takes a timely look at the role global fits have played in physics developments throughout the LHC era, how that role has also been shaped and changed by those developments and the big role they can play in the future. The rest of this review is structured as follows. In \cref{sec:physics}, we further review past and present applications of global fits. A discussion of future applications, prospects, and challenges will be summarized in \cref{sec:future}. We draw our conclusions in \cref{sec:summary}.

\section{Past and present physics applications}\label{sec:physics}

Before the LHC era, in the early 1990s, global fits were used in precision tests of the electroweak sector of the SM~\cite{Altarelli:1997et,Altarelli:2004fq,Erler:2019hds}. As described in \cref{sec:ewpts}, results from these fits included successful predictions for the top mass, the Higgs mass, and ruling out specific strongly-interacting models of electroweak symmetry breaking. In the mid 1990s and early 2000s global fits were expanded to specific extensions of the SM that could include e.g., direct searches at colliders and DM observables. The first fits were spurred by anomalies found in precision measurements, e.g., a global fit of a minimal supersymmetric model in light of tension in measurements of the strong coupling constant from LEP~\cite{Kane:1995nw}, and in light of the anomalous magnetic moment of the muon and the branching ratio for the process $b\to s \gamma$~\cite{deBoer:2001nu}. Around the same time, global fits in cosmology were maturing~\cite{Lewis:2002ah}, including the public computer software \code{CosmoMC}~\cite{cosmomc}. These fits used Markov Chain Monte Carlo (MCMC) algorithms as the impact of the MCMC revolution in the early 1990s in statistics~\cite{Robert2011,f47e9344-75dc-3782-a811-9e51a952e8bf} arrived. These developments influenced particle physics, leading to several global fits of constrained supersymmetric models using MCMC~\cite{Baltz:2004aw,Allanach:2005kz,RuizdeAustri:2006iwb}, random sampling~\cite{Profumo:2004at}, optimization methods~\cite{Buchmueller:2007zk,Buchmueller:2008qe} and grids~\cite{Baer:2003yh}. Possibly inspired by cosmology and MCMC, global fits were at first mostly Bayesian, but later included frequentist analyses~\cite{Buchmueller:2007zk,Buchmueller:2008qe}. These fits were made possible by an ecosystem of scientific software for computing predictions, including e.g., \code{SOFTSUSY}~\cite{Allanach:2001kg},
\code{micrOMEGAs}~\cite{Belanger:2001fz,Belanger:2006is,Alguero:2023zol},
\code{DarkSUSY}~\cite{Gondolo:2000ee,Bringmann:2018lay}, \code{HDECAY}~\cite{Djouadi:1997yw} and \code{FeynHiggs}~\cite{Heinemeyer:1998yj}, that communicated using the SLHA convention~\cite{Skands:2003cj}. These tools were combined with algorithms for statistical computation by small teams of researchers in packages such as \code{SFitter}~\cite{Lafaye:2004cn,Lafaye:2007vs}, \code{SuperBayeS}~\cite{2011ascl.soft09007R}, \code{Fittino}~\cite{Bechtle:2004pc} and \code{MasterCode}~\cite{mastercode}.

The first wave of global fits pioneered the methodology and toolchains, and mapped out the parameter spaces for the first time. A second wave of global fits targeted the start of the LHC, and forecasted discovery prospects for the Higgs and supersymmetric particles there~\cite{Roszkowski:2007fd,Allanach:2007qk,Buchmueller:2007zk,Buchmueller:2008qe,Cabrera:2009dm,Bechtle:2010igv}
and at the Tevatron~\cite{Roszkowski:2006mi}. The forecasts in \refscite{Allanach:2007qk,Cabrera:2009dm} took into account  electroweak fine-tuning through a fine-tuning penalty found to be automatically present in Bayesian analyses. These global fits were updated in 2011 immediately after the first $35/\text{pb}$~\cite{CMS:2011xek} of collision data from the  LHC at $7\tev$~\cite{Fowlie:2011mb,Allanach:2011ut,Allanach:2011wi,Buchmueller:2011aa,Strege:2011pk,Bechtle:2011dm}. Incorporating the LHC results precisely required expensive MC simulation of events --- this involves hard-scattering processes, resonance decays, initial and final-state radiation, hadronization, hadron decays and the modelling of the detector response (see e.g., \refcite{Buckley:2011ms} for a review) --- and was the first major challenge posed in the LHC era. This issue was tackled in \refscite{Fowlie:2011mb,Allanach:2011ut,Allanach:2011wi} by precomputing expensive calculations on a two-dimensional grid
and neglecting or approximating the dependence on other parameters.
\Refscite{Strege:2011pk,Buchmueller:2011aa} also approximated the parameter space as two-dimensional and used an approximate formula based on the published exclusion limits on a two-dimensional plane.  Lastly, \refcite{Bechtle:2011dm} performed simulation at every point. These approaches do not generalize well to models where more than two parameters are relevant.

The Higgs discovery in 2012~\cite{ATLAS:2012yve,CMS:2012qbp} was a turning point in global fitting trends. There were two major implications for updated fits in constrained supersymmetric models~\cite{Buchmueller:2011ab,Fowlie:2012im,Buchmueller:2012hv,Balazs:2012vjy,Ellis:2012rx,Bechtle:2012zk,Strege:2012bt}. First, the relatively heavy mass of $125\gev$ was hard to accommodate in constrained models, and pushed superpartner masses towards the multi-TeV region, somewhat at tension with naturalness, a major motivation for SUSY. At this moment, there was an increased interest in relaxed, non-minimal and non-supersymmetric models.  Second, the multi-TeV region posed challenges for precise computations of the Higgs mass and other observables, due to poor convergence of perturbative calculations involving large logarithms~\see{Barbieri:1990ja,Okada:1990gg,Haber:1993an,PardoVega:2015eno,Bahl:2016brp,Athron:2016fuq,Slavich:2020zjv,Kwasnitza:2025mge}.
The changes in global fits for a simple SUSY model over the last twenty years are illustrated in \cref{fig:m0m12}, which shows results for two soft-breaking mass parameters. The results from early fits, e.g.~\cref{fig:m0m12-a,fig:m0m12-b}, were presented as two-dimensional slices of parameter space. As discussed, closer to the launch of the LHC more advanced multi-dimensional sampling methods were introduced. E.g.~\cref{fig:m0m12-c,fig:m0m12-d} show marginalized posterior pdfs from Bayesian fits
and \cref{fig:m0m12-e,fig:m0m12-f}
show confidence intervals based on the profile likelihood from frequentist fits.
The preferred regions in these plots (continuing to \cref{fig:m0m12-g,fig:m0m12-h,fig:m0m12-i}) were impacted by new experimental data as well as by improvements in the sampling algorithms and updates in the theoretical calculations.  The physics implications reflected by the changes in these plots will be discussed in \cref{Sec:SUSY}.

After the Higgs discovery, the Higgs properties were measured and found to be consistent with the SM, and more and more null results in ATLAS and CMS searches for new physics flowed in from runs I and II.  The shift in interest continued, with a new focus on bottom-up models, including phenomenological models, such as MSSM-type models with parameters defined at the weak scale (see \cref{Sec:SUSY}) and effective field theories (see \cref{sec:eft}). In this period, CMS and ATLAS conducted an increasingly broad and sophisticated set of searches for new physics, using complex final states and cuts, including ML classifiers. Incorporating these analyses into a global fit requires detailed modelling, including event generation, parton showering, detector simulation, and emulating cutflows. For example, a recent analysis of neutralinos and charginos combined 27 relevant ATLAS and CMS searches~\cite{GAMBIT:2023yih} and there are many more searches for other supersymmetric particles~\cite{Canepa:2019hph}.

As a consequence of these computational challenges --- more models with more parameters, large logarithms and MC event generation --- and the time demands of implementing and validating so many searches,
global fits of extensions to the SM are now a challenging undertaking, typically performed by teams or collaborations of specialists with access to high-performance computing resources.
There has been significant progress in the development of public tools dedicated to global fits that are suitable for all challenges. The Global and Modular BSM Inference Tool Collaboration (\gambit{}) develops public software for intensive global fitting of a range of BSM models~\cite{GAMBIT:2017yxo} and uses it to perform global fits of supersymmetric models~\cite{GAMBIT:2017zdo,GAMBIT:2017snp,GAMBIT:2018gjo,GAMBIT:2023yih}, DM models~\cite{GAMBIT:2017gge,Athron:2018ipf,GAMBIT:2018eea,GAMBIT:2021rlp,Chang:2022jgo,Chang:2023cki}, neutrino physics~\cite{Chrzaszcz:2019inj}, axions and axion-like particles~\cite{Hoof:2018ieb,Athron:2020maw,Balazs:2022tjl}, two Higgs doublet models~\cite{Athron:2021auq,Beniwal:2022kyv,Athron:2024rir} and even cosmological models through \code{CosmoBit}~\cite{GAMBITCosmologyWorkgroup:2020rmf}.
As we discuss in \cref{sec:neutrinos}, the latter connects cosmology and particle physics, allowing one to fully explore e.g., neutrino mass or decaying DM.
In this period, \code{EasyScan\_HEP}~\cite{Shang:2023gfy},  \code{xBit}~\cite{Staub:2019xhl} and \code{HepFit}~\cite{DeBlas:2019ehy} were also developed.

\gambit{}, through \code{ColliderBit}~\cite{GAMBIT:2017qxg}, incorporates LHC searches by performing event simulation at every parameter point, and measurements through \code{Rivet}~\cite{Buckley:2010ar} and \code{Contur}~\cite{Butterworth:2016sqg}. A similar strategy is taken by recasting tools \code{CheckMate}~\cite{Drees:2013wra,Dercks:2016npn} and \code{MadAnalysis}~\cite{Conte:2012fm}. \code{SModelS}~\cite{Kraml:2013mwa,Altakach:2024jwk} provides a computationally cheaper alternative, based on mapping new physics scenarios to simplified models. In any case, a trend towards public likelihoods~\cite{Cranmer:2021urp,LHCReinterpretationForum:2020xtr} and likelihood preservation~\cite{Heinrich:2021gyp,Feickert:2020wrx,ATLAS:2019oik}, sometimes using ML~\cite{CMS:2025cwy,Coccaro:2019lgs}, could assist in the implementation of LHC search results for global fits.

New models could mean that expressions for model predictions must be computed and coded by hand, which is both time-consuming and error-prone. This challenge is overcome by automated tools. Using computer algebra systems, automated tools can generate symbolic expressions for the Feynman rules of new physics models, renormalization group equations and matrix elements. They can, furthermore, then write  numerical routines that can be interfaced with existing codebases for the mass spectrum, one-loop form factors, new particle decay branching ratios and cross sections.
Modern examples of such algebraic codes include~\code{LanHep}~\cite{Semenov:1996es,Semenov:1998eb,Semenov:2008jy,Semenov:2014rea},
\code{CalcHep}~\cite{Belyaev:2012qa}, \code{CompHep}~\cite{CompHEP:2004qpa},
\code{FeynRules}~\cite{Christensen:2008py,Alloul:2013bka},
\code{SARAH}~\cite{Staub:2010jh,Staub:2012pb,Staub:2013tta,Goodsell:2015ira},
and \code{MARTY}~\cite{Uhlrich:2020ltd}.
Programs such as
\code{FlexibleSUSY}~\cite{Athron:2014yba,Athron:2017fvs,Athron:2021kve},
\code{Spheno}~\cite{Porod:2003um,Porod:2011nf} and \code{micrOMEGAs}~\cite{Belanger:2001fz,Belanger:2006is,Alguero:2023zol} use such packages to create completely new software for predicting observables in arbitrary models specified by the user.
In the context of global fits, \code{GUM}~\cite{Bloor:2021gtp} and \code{XBit}~\cite{Staub:2019xhl} take advantage of these tools to generate numerical code for a global fit for a new physics model defined only in symbolic form.

Thus, at present there are tools to tackle these problems and global fits are being used in a broad range of supersymmetric and non-supersymmetric models.  In the remainder of this section, we zoom in on the impact of global fits in several particular areas of new physics.

\subsection{Electroweak precision tests}\label{sec:ewpts}

Electroweak precision tests (EWPTs) are precise measurements of electroweak observables from LEP, SLC, Tevatron, and the LHC that constrain the electroweak sector of the SM~\cite{Altarelli:1997et,Altarelli:2004fq,Erler:2019hds,deBlas:2021wap}. The key measured observables include e.g., the mass of the $W$ boson, the effective Weinberg angle ($\sin^2\theta_{W}$), various forward-backward asymmetries ($A_{\rm FB}$), and the total and partial decay widths of the $Z$ boson, including its invisible decay width. As emphasized in \refcite{Hollik:1988ii}, precise calculations of radiative corrections were required to match the precise measurements at LEP and SLC.
The combination of precise measurements and theoretical predictions allowed global fits to constrain fundamental SM parameters that were at the time unknown, such as the top quark mass ($m_t$) and even the Higgs mass ($m_H$).

For example, \refcite{Langacker:1991an} showed that the combination of electroweak precision observables led to constraints on the top quark mass: $m_t \lesssim 182\gev$ at $95\%$ confidence well before its direct discovery at the Tevatron~\cite{CDF:1995wbb,D0:1995jca} and compatible with the current value from direct measurements: $m_t = 172.57 \pm 0.29\gev$~\cite{ParticleDataGroup:2024cfk}.
Similarly, before the Higgs discovery at the LHC,
\refcite{Baak:2011ze} predicted the Higgs boson mass to be $120^{+12}_{-5}\gev$ using electroweak data and bounds from LEP.
This prediction is consistent with the current value of the observed Higgs boson mass, $m_H = 125.20 \pm 0.11\gev$~\cite{ParticleDataGroup:2024cfk}.
These remarkable predictions for SM parameters could not have been made without the use of global fits. Besides the electroweak sector, global fits are now a standard technique for determining SM parameters in the neutrino sector~(see \cref{sec:neutrinos}),
the Cabibbo-Kobayashi-Maskawa (CKM) mixing matrix~\cite{Charles:2004jd,Charles:2015gya} and parton distribution functions~\cite{NNPDF:2017mvq,Hou:2019efy}.

As proposed by Peskin and Takeuchi in 1991~\cite{Peskin:1991sw}, EWPTs can also be used to constrain BSM physics in a model-independent manner using the so-called \emph{oblique parameters}, $S$, $T$ and $U$. The oblique parameters parameterize new contributions to electroweak precision observables using radiative corrections to the universal two-point functions for the SM gauge bosons. Including $S$ and $T$ in a global fit but assuming $U=0$, hadron collider and electroweak precision data requires $S=-0.05 \pm 0.07$ and $T=0.00 \pm 0.06$ with a positive 93\% correlation between the two parameters~\cite{ParticleDataGroup:2024cfk}.
Thus, many models of new physics that predict $U=0$ can be tested by computing their contributions to $S$ and $T$.\footnote{For models that can predict significant contributions to $U$ three parameter fits outputting $S$, $T$ and $U$ can be used.}   For example, in the early 1990s technicolor models were strongly constrained by their predicted contributions to the $S$ parameter~\cite{Peskin:1990zt}. However, the oblique parameters are limited; they cannot, e.g., parameterize new physics that couples directly to fermions; these effects can be captured in effective field theory described in \cref{sec:eft}.

\subsection{Supersymmetric models}\label{Sec:SUSY}

\newcommand{\addfig}[5]{%
	\begin{subfigure}[b]{#4\linewidth}
		\centering
		\includegraphics[height=#3]{#2}
		\caption{#1}
		\label{fig:m0m12-#5}
	\end{subfigure}%
}

\begin{figure}
	\centering
	\addfig{\textbf{2003}. Fig.~3 of \refcite{Baer:2003yh}}{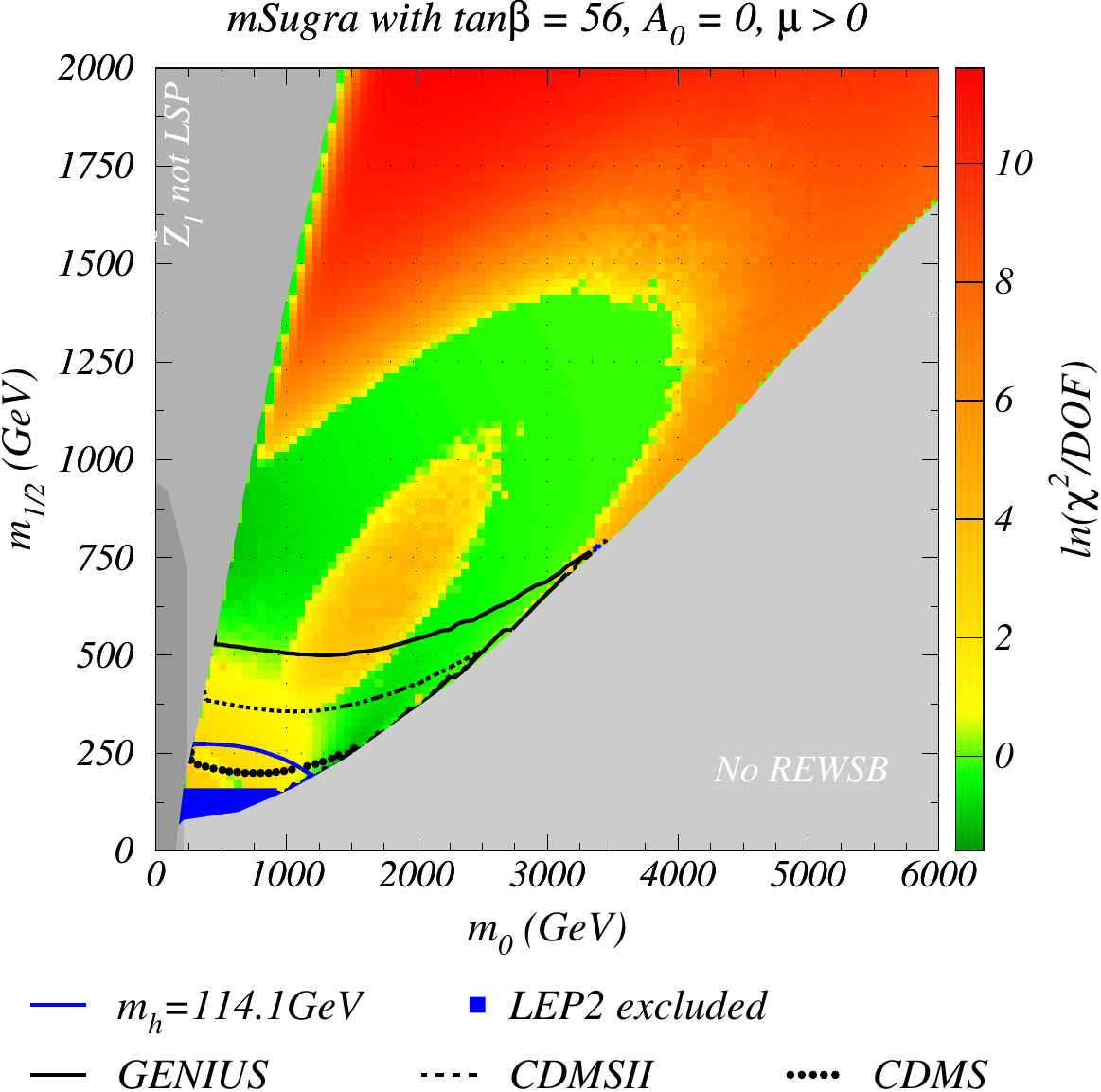}{5.25cm}{0.33}{a}%
	\addfig{\textbf{2003}. Fig.~6 of \refcite{deBoer:2003xm}}{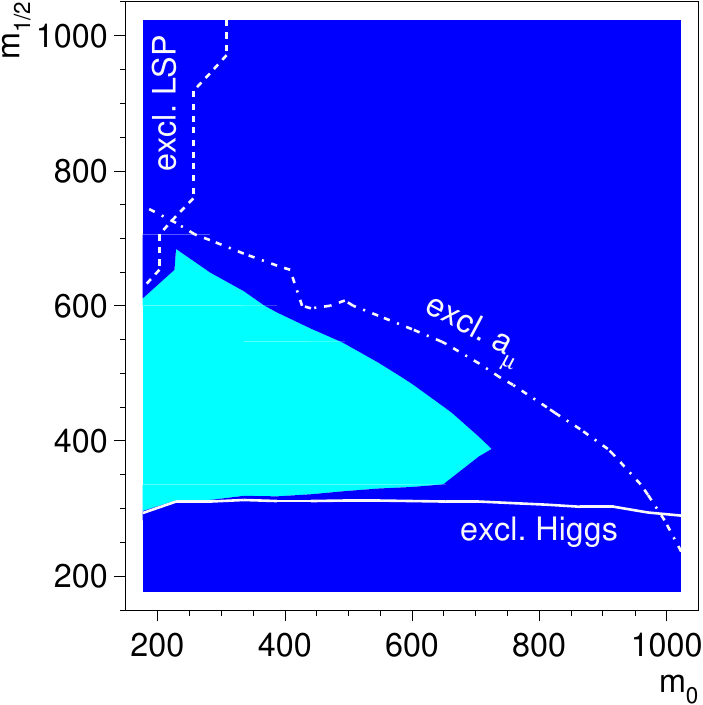}{5.25cm}{0.33}{b}%
	\addfig{\textbf{2006}. Fig.~3 of \refcite{RuizdeAustri:2006iwb}}{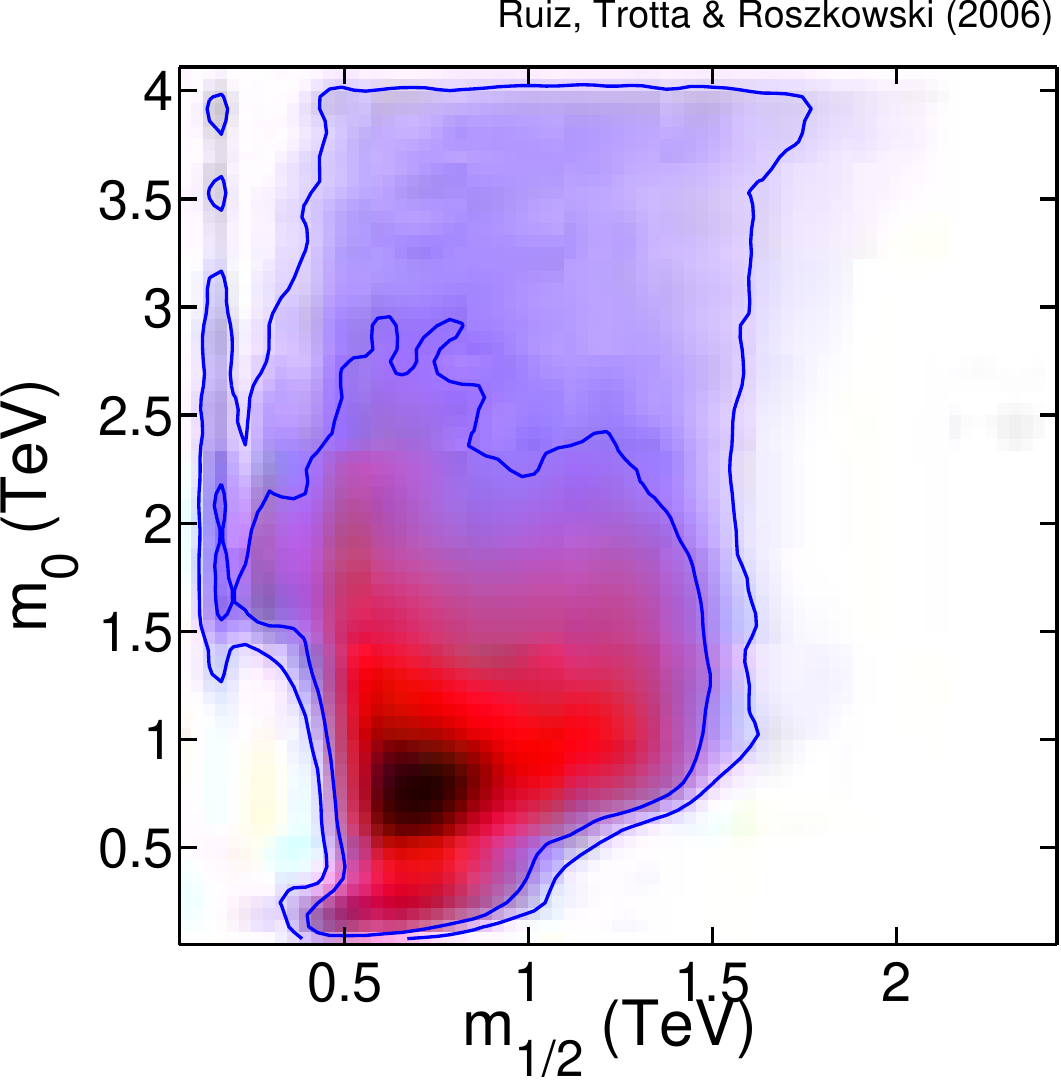}{5.25cm}{0.33}{c}%

	\bigskip

	\addfig{\textbf{2007}. Fig.~2 of \refcite{Allanach:2007qk}}{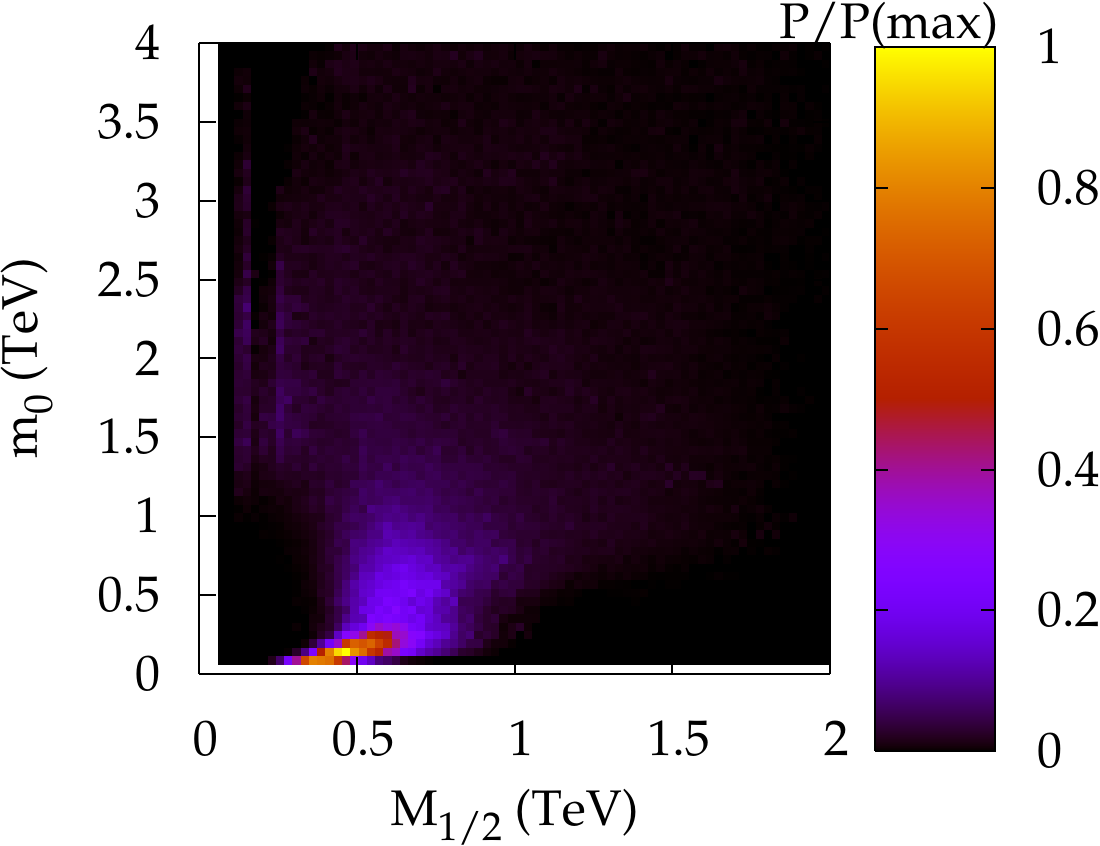}{4cm}{0.325}{d}%
	\addfig{\textbf{2009}. Fig.~2 of \refcite{Bechtle:2010igv}}{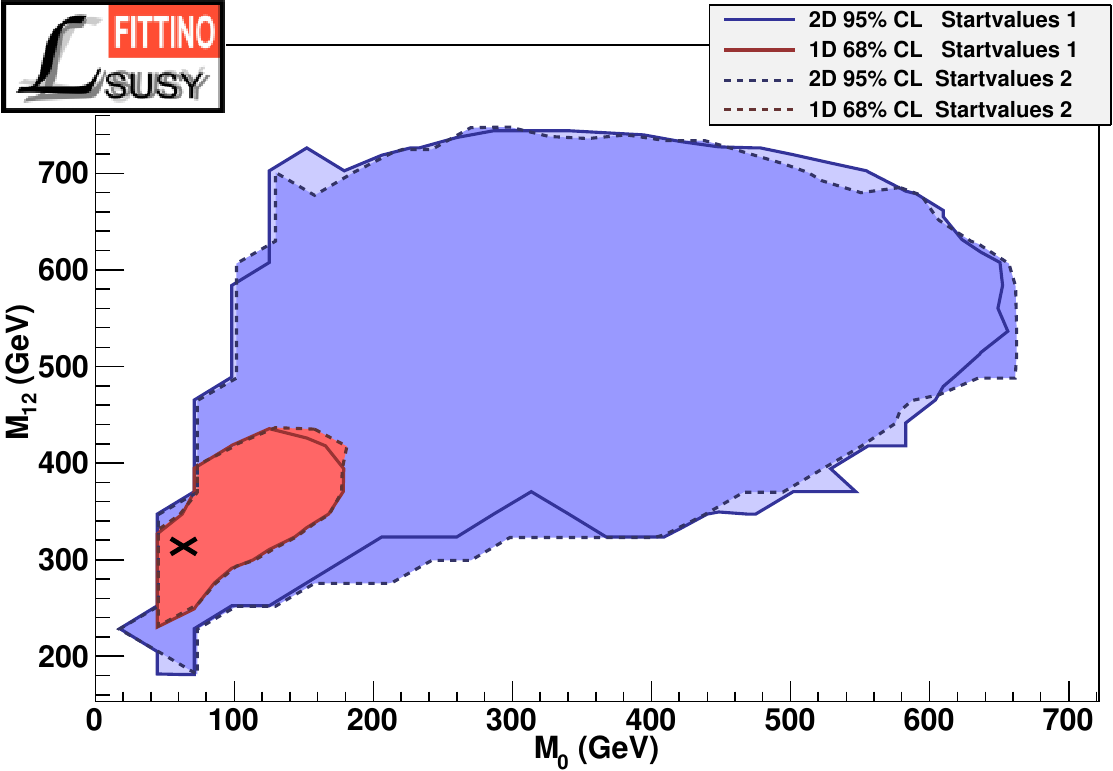}{4cm}{0.35}{e}%
	\addfig{\textbf{2011}. Fig.~1 of \refcite{Buchmueller:2011aa}}{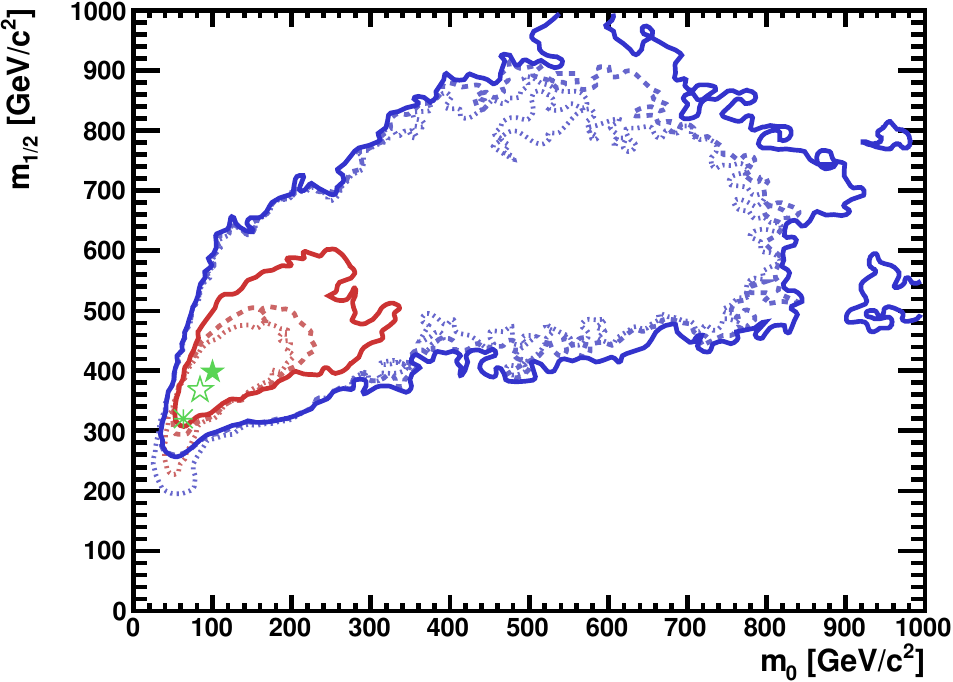}{4cm}{0.325}{f}%

	\bigskip

	\addfig{\textbf{2012}. Fig.~2 of \refcite{Fowlie:2012im}}{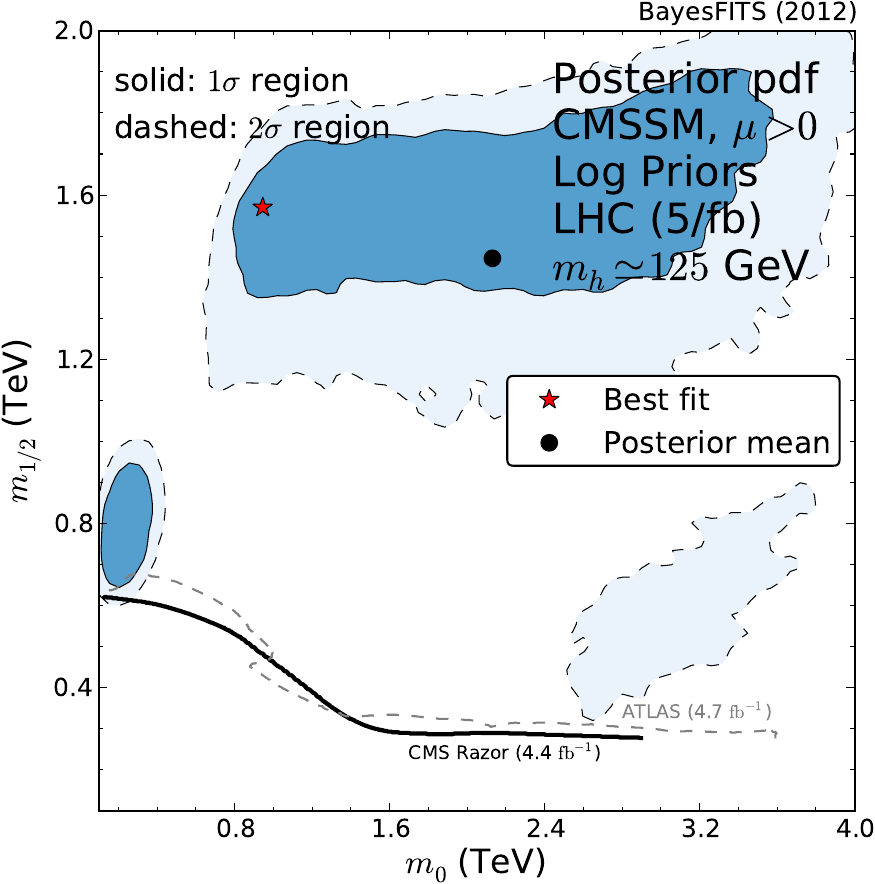}{4.35cm}{0.28}{g}%
	\addfig{\textbf{2015}. Fig.~1 of \refcite{Buchmueller:2015uqa}}{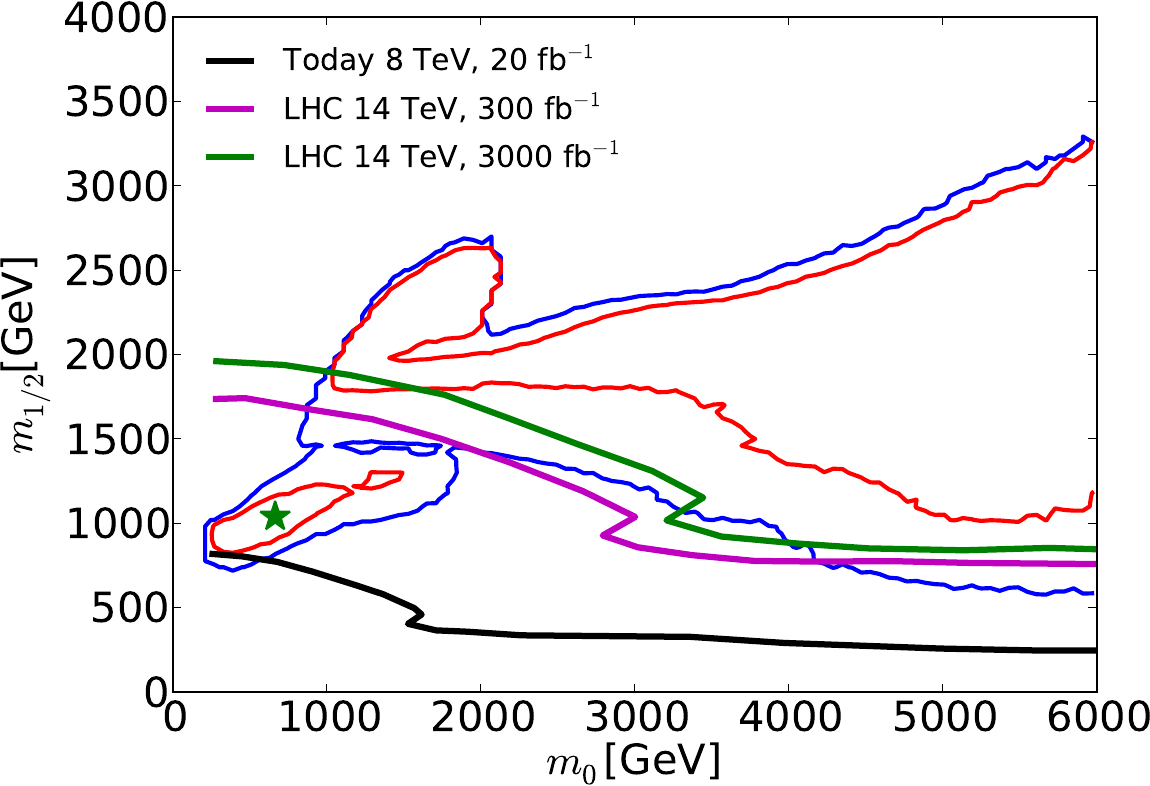}{4.35cm}{0.37}{h}%
	\addfig{\textbf{2017}. Fig.~2 of \refcite{GAMBIT:2017snp}}{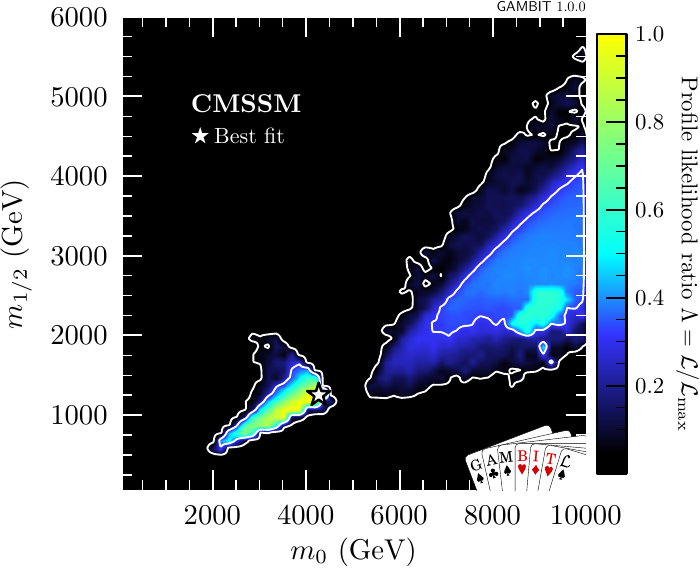}{4.35cm}{0.35}{i}%
	\caption{Evolution of global fits of the CMSSM in ($m_0$, $m_{1/2}$). The fits use different methodologies and datasets.}
	\label{fig:m0m12}
\end{figure}

Supersymmetric (SUSY) extensions of the SM \seeone{Martin:1997ns} are appealing because they can address many outstanding problems of the SM, such as the hierarchy problem \seeone{Peskin:2025lsg} and astrophysical observations of DM (see e.g.~\refscite{Feng:2022rxt,Cirelli:2024ssz,Balazs:2024uyj}).
Most phenomenological exploration of SUSY takes the Minimal Supersymmetric Standard Model (MSSM) as a starting point because it supersymmetrizes the SM with minimal field content and is agnostic about the mechanism of soft SUSY breaking at high-energy scales.
The latter means that it includes all gauge-invariant terms that softly break SUSY. A challenge to this approach is that the parameter space of the MSSM then contains more than one hundred parameters. This leads to practical difficulties for phenomenological studies of the MSSM in terms of handling so many parameters.
To improve this situation, one must make assumptions about the soft-breaking parameters. Models can be divided into high-scale and weak-scale models depending on the nature of these assumptions.

In high-scale models, one fixes the full set of soft-breaking SUSY parameters in terms of a much smaller set of parameters using constraints inspired by theories about SUSY breaking and/or by embedding the MSSM into grand unified theories (GUTs).  These constraints are typically applied  significantly above the electroweak scale, e.g.\ at the scale where the gauge couplings unify, $M_\text{GUT} \sim10^{16}\gev$.
The constrained MSSM (CMSSM)~\cite{Nilles:1983ge,Kane:1993td} is the most-studied SUSY model in this category. The number of parameters in the CMSSM is reduced to four and a half: the soft-breaking scalar mass ($m_0$), the unified soft-breaking gaugino mass ($m_{1/2}$), the ratio of the vacuum expectation values of the two scalar Higgs doublets ($\tan\beta \equiv v_u / v_d$), the unified trilinear coupling ($A_0$) for scalars and the sign of the Higgsino mass parameter ($\sign\mu$). Global fits of the CMSSM have been extensively performed in the literature; see e.g.~\refscite{Henrot-Versille:2013yma,Martinez:2009jh,Bechtle:2013mda,Buchmueller:2013psa,Bechtle:2014yna,Roszkowski:2014wqa,Fowlie:2014awa,Buchmueller:2013rsa,Han:2016gvr}. The most recent global fit of the CMSSM has been performed in \refcite{GAMBIT:2017snp} which uses measurements and bounds from dark-matter searches, flavour and electroweak observables, LEP searches, and results from LHC Run-I and (early) Run-II that were available at the time.
As well as the CMSSM, high-scale models motivated by anomaly mediated supersymmetry breaking~\cite{Bagnaschi:2016xfg} and $SU(5)$ GUTs~\cite{Bagnaschi:2016afc} have been considered, as well as models where the boundary conditions are relaxed to allow non-universal Higgs masses~\cite{Buchmueller:2012hv,Strege:2012bt,Buchmueller:2013rsa,Roszkowski:2014wqa,GAMBIT:2017snp}, or the scale they are applied at is varied below the GUT scale~\cite{Costa:2017gup}.

\Cref{fig:m0m12} shows a collection of results on the ($m_0$, $m_{1/2}$) plane of the CMSSM over the last twenty years.  The changes in these plots reflect advances and differences in methods for performing global fits, advances in theoretical calculations, and new experimental data.  Both \cref{fig:m0m12-a,fig:m0m12-b} show results from the early 2000s, when global fits were in their infancy.\footnote{The shapes of the allowed regions in \cref{fig:m0m12-a,fig:m0m12-b} are dramatically different because \cref{fig:m0m12-b} included early data from the BNL experiment for muon $g-2$~\cite{Muong-2:2002wip}, which showed a deviation from the SM.
Satisfying this deviation required light sparticles and thus led to an upper limit on the soft-breaking masses.} \Cref{fig:m0m12-c,fig:m0m12-d,fig:m0m12-e} show pre-LHC forecasts that used more sophisticated Bayesian and frequentist statistical methods. In the subsequent post-LHC plots, \cref{fig:m0m12-f,fig:m0m12-g,fig:m0m12-h,fig:m0m12-i}, the soft-breaking masses were pushed further and further upwards by null results of searches for sparticles at the LHC and the Higgs mass.

These fits delineated the viable regions of the CMSSM parameter space and are critical for designing tests to further exclude it. For example, in the context of DM, the first wave of global fits identified distinct viable DM annihilation mechanisms that might be in reach of future collider and direct detection searches, including stau co-annihilation~\cite{Ellis:1999mm}.
Post-LHC run I global fits showed that stau co-annihilation was ruled out by direct searches for gluinos at the LHC in combination with other data~\cite{Bagnaschi:2015eha,Han:2016gvr,GAMBIT:2017snp}. The improved optimization algorithms employed in \refcite{GAMBIT:2017snp} subsequently identified a stop co-annihilation mechanism for explaining DM that was not excluded by searches at the LHC or other data. This stop co-annihilation mechanism allowed for sparticle masses that were lighter than those in any other region.

\emph{Weak-scale models}, on other hand, focus on weak-scale soft-breaking parameters that most directly affect experimental observables. In contrast to high-scale models, they are agnostic about the mechanism of SUSY breaking and grand unification. The weak-scale MSSM is often referred to as the phenomenological MSSM (pMSSM)~\cite{MSSMWorkingGroup:1998fiq}. Specific choices of parameter relations reduce the number of free parameters in the pMSSM and models are often denoted by pMSSM$n$, where $n$ denotes the number of free parameters.

Global fits of a pMSSM$7$, pMSSM$10$, pMSSM$11$, pMSSM$15$ and pMSSM$20$ have been carried out~\cite{AbdusSalam:2009qd, AbdusSalam:2012ir, AbdusSalam_2014, Strege_2014,  Bertone:2015tza, Bagnaschi:2015eha, AbdusSalam:2015uba, deVries:2015hva, Kowalska:2016ent, Bagnaschi:2017tru, GAMBIT:2017zdo}.
These pMSSM$n$ fits suffer from the curse of dimensionality, with computational cost that can be exponential in dimension $n$. Although sampling algorithms and computing power have improved over the LHC era, simulating LHC searches is so expensive that these global fits are arguably harder than they were prior to the LHC.  Before the LHC, Bayesian fits of the pMSSM20 were carried out in \refcite{AbdusSalam:2009qd}. These were later updated with LHC 7\tev results~\cite{AbdusSalam:2012ir}, and the Higgs discovery~\cite{AbdusSalam_2014},  demonstrating that high-dimensional explorations were feasible, but only with approximate collider constraints. With LHC Run~I 8\tev data, a pMSSM15 study~\cite{Strege_2014} (2014) carried out one of the first statistically convergent frequentist global fits in a high-dimensional setting, implementing detailed likelihoods for ATLAS 0-lepton and 3-lepton searches through recast simulations performed by post-processing selected subsets of the sampled points. The pMSSM7 study~\cite{GAMBIT:2017zdo} (2017) maintained the frequentist framework. Compared to~\cite{Strege_2014}, this study considered a significantly lower-dimensional parameter space, but incorporated simulations of 13 higher-luminosity ATLAS and CMS searches from both Run I and Run II, leading to a significantly larger per-point simulation cost. Finally, the pMSSM11 study~\cite{Bagnaschi:2017tru} (also 2017) opted for a different balance between parameter space dimensionality and per-point computational cost by implementing Run~I and Run~II LHC constraints through the much faster but more approximate \code{FastLim}~\cite{Papucci:2014rja} approach (similar in spirit to the \code{SModelS}~\cite{Kraml:2013mwa,Altakach:2024jwk} approach), which applies conservative LHC limits without per-point simulations.  In short, the Bayesian pMSSM20 fits represented the earliest broad explorations, while the pMSSM15, pMSSM7 and pMSSM11 stand out as technically more challenging frequentist analyses, each settling for a different trade-off between model generality and the computational cost of simulations and parameter space exploration. Overall, one of the main messages from these studies was that LHC limits on sparticle masses were significantly weaker in the pMSSM than in more restrictive SUSY frameworks, such as constrained or simplified scenarios.

The so-called electroweak-MSSM (EWMSSM) is an extreme example of a weak-scale model that reduces the number of free parameters in the pMSSM. The EWMSSM considers only the chargino and neutralino sectors of the pMSSM; the other states are decoupled, as perhaps hinted at by the null results from searches for sleptons and squarks at the LHC and the Higgs boson mass. Thus, this  model has only four parameters: the electroweak gaugino masses ($M_1$ and $M_2$), the Higgsino mass parameter ($\mu$) and $\tan\beta$. The splitting of electroweakinos from other sparticles could be described as a split-SUSY scenario~\cite{Giudice:2004tc,Arkani-Hamed:2004zhs}.
\Refcite{GAMBIT:2018gjo} performed a global fit of this model in light of LHC searches for neutralinos and charginos at the LHC and LEP.
The global fit revealed that no part of the mass plane of the lightest neutralino and lightest chargino is completely forbidden, as LHC and LEP searches could be evaded by adjusting parameters orthogonal to that plane.  This could only be revealed by performing a global fit using the profile likelihood and has provided impetus for new experimental searches. In \refcite{GAMBIT:2023yih} this global fit was extended by adding an approximately massless gravitino. Whilst the gravitino collider signatures lead to strong constraints on the masses of the electroweakinos, viable scenarios with electroweakinos as light as $140\gev$ were identified. In addition to direct searches for electroweakinos, measurements of SM-like final states, applied using \code{Contur}~\cite{CONTUR:2021qmv}, were found to be important; exploring the complementarity between direct and indirect constraints was possible in a global fit. Finally, electroweakino models were used in global fits by the ATLAS collaboration; \refcite{ATLAS:2016} performed a global fit of a weak scale model of the neutralinos, charginos and a pseudo-scalar Higgs, finding viable points with electroweakinos lighter than $65\gev$.

Going beyond the minimal particle content, one can consider the Next-to-Minimal Supersymmetric Standard Model (NMSSM; see e.g., \refcite{Ellwanger:2009dp}).
The NMSSM extends the MSSM by a new gauge singlet fermion and its superpartners, two scalar singlets, which can in principle have a significant impact on the phenomenology. Through global fits, \refscite{Balazs:2009su,Lopez-Fogliani:2009qdp,Kowalska:2012gs} found that the semi-constrained NMSSM (scNMSSM) was a compelling alternative to the CMSSM, offering interesting and distinct DM scenarios, though the fits ultimately preferred rather CMSSM-like scenarios and disfavoured scenarios where the SM-like Higgs was not the lightest Higgs.\footnote{Fully-constrained versions of the NMSSM with fully universal soft scalar and trilinears are technically challenging to solve in a convergent iteration solving the two scale boundary value problem and are in any case highly constrained, presenting a challenge to obtain a $125\gev$ SM-like Higgs~\cite{Djouadi:2008yj,Arbey:2011ab}. As a result, the semi-constrained NMSSM allows a non-universal soft scalar singlet which also has some theoretical motivation.}
On the other hand, mostly as a result of a Bayesian penalty for fine-tuning required by results from the LHC and the $\mu$-problem, \refcite{Fowlie:2014faa} found that the scNMSSM was favoured and less fine-tuned than its simpler counterpart, the CMSSM.
Significant phenomenological differences can occur when relaxing the assumptions of the scNMSSM or considering weak scale NMSSM models. For example, a global fit of a 26 parameter phenomenological NMSSM found preference for an additional Higgs below the mass of the observed one~\cite{AbdusSalam:2017uzr}. However, to date few detailed global fits of such scenarios have been carried out.
Finally, dramatically different phenomenology has been demonstrated in other non-minimal supersymmetric extensions of the SM, including $R$-parity violating scenarios~\cite{Mohapatra:2015fua}; extended gauge groups, such as $E_6$-inspired models with an extra $U(1)$ that solves the $\mu$-problem where exotics are present~\cite{Langacker:2008yv,King:2005jy}; extra supermultiplets, such as the $\mu\nu$SSM~\cite{Lopez-Fogliani:2005vcg} that solves the $\mu$-problem using right-handed neutrinos, $\nu$; and $R$-symmetric models that require Dirac gauginos~\cite{Kribs:2007ac}.
However, although the parameter spaces of such models have been explored~(e.g.,~\refscite{Athron:2016qqb,Kpatcha:2019qsz}), to date no statistical global fits have been performed.

\subsection{Two Higgs doublet models}\label{sec:thdm}

In light of the Higgs discovery and the absence of SUSY at the LHC, there was renewed interest in two Higgs doublet models (2HDMs) as simple and obvious extensions to the SM, as well as the scalar singlet model and Higgs portal models discussed in \cref{sec:dm}. The second Higgs doublet leads to a richer scalar sector containing eight real degrees of freedom. After electroweak symmetry breaking, there are two \CP-even scalars ($h^0$ and $H^0$), one \CP-odd scalar ($A^0$) and a pair of charged scalars ($H^\pm$). There are non-trivial implications for electroweak symmetry breaking, flavour physics, electroweak baryogenesis, and collider phenomenology (see \refcite{Branco:2011iw} for a comprehensive review). There are several versions of the 2HDM, distinguished by the symmetries of the scalar potential and the vacuum. The simplest 2HDMs possess a softly-broken $\mathbb{Z}_2$ symmetry~\cite{Glashow:1976nt} and their parameters are real to avoid large tree-level flavour changing neutral currents (FCNCs). This leads to four realizations of the 2HDM --- known as flavour-conserving models and usually denoted as type I, type II, type X and type Y:
\begin{enumerate}
	\item \textbf{Type I}: Only one Higgs doublet, $\Phi_{2}$, couples to all fermions.
	\item \textbf{Type II}: The Higgs doublet $\Phi_{2}$ couples to up-type quarks and $\Phi_{1}$ couples to down-type quarks and charged leptons. The Higgs sector coincides with that of the MSSM.
	\item \textbf{Type X}: Similar to type I but leptons couple to $\Phi_{1}$.
	\item \textbf{Type Y}: Similar to type II but leptons couple to $\Phi_{2}$.
\end{enumerate}
These models have noticeably different phenomenological implications (see \refcite{Aoki:2009ha} for a comprehensive analysis of these models).

Comprehensive fits of these models~\cite{Mahmoudi:2009zx,Aoki:2009ha,Basso:2012st,Celis:2013rcs,Keus:2015hva,Muhlleitner:2017dkd,Haller:2018nnx,Arbey:2017gmh,Misiak:2017bgg,Li:2024kpd,Atkinson:2021eox,Atkinson:2022pcn} have combined constraints from electroweak precision data, Higgs boson coupling measurements, flavour observables, and direct collider searches. This combination of data allows them to place lower limits on the mass of the charged Higgs boson. For example
in the type II model
\refscite{Arbey:2017gmh,Misiak:2017bgg,Li:2024kpd} find that
the charged Higgs mass must be greater than approximately $600\gev$ at the 95\% confidence level. By using an updated NNLO prediction of $\bar B\to X_s\gamma$~\cite{Misiak:2020vlo}, \refcite{Atkinson:2021eox}
finds a stronger bound of $860\gev$ at 95\%. In contrast, the fits find no lower limit for the charged Higgs mass in the type I and X models. Additionally, the fit in \refcite{Beniwal:2022kyv} generated a strong constraint on the mass splittings between BSM Higgs bosons, which strongly influences the strength of the electroweak phase transition in the 2HDM~\cite{Song:2022xts}. In all these fits small deviations from the alignment limit were found, implying the possibility of discriminating these models from the SM using precision Higgs coupling measurements at future colliders~\cite{Gu:2017ckc}.
Global fits~\cite{Chowdhury:2015yja} also reveal that vacuum stability and naturalness can place strong restrictions on the parameter space. Requiring vacuum stability at the EW scale in addition to applying collider constraints severely restricts deviations from alignment, and in type II and type Y models the global fit puts indirect lower bounds on the neutral scalar masses.  Both of these constraints tighten further when vacuum stability is imposed at the Planck scale.   Similarly, imposing naturalness principles can constrain the mass scale from above since the 2HDM is not protected from large quadratic corrections to the Higgs mass, and avoiding these requires a UV completion of the 2HDM.  Thus, using naturalness, \refcite{Chowdhury:2015yja} place an upper bound on the mass scale of such UV completions of $5\tev$ for type I and X models and $3.7\tev$ for type II and Y models. The latter is stronger due to the stricter constraints on the parameter space in these models.

The flavour-aligned 2HDM is a more general framework for avoiding FCNC, that does not use a $\mathbb{Z}_2$ symmetry, unlike the four types discussed above. In the flavour-aligned 2HDM, flavour conservation in the neutral Higgs couplings is imposed by requiring flavour alignment in the Yukawa sector~\cite{Pich:2009sp}. Three complex parameters relate the up, down and lepton Yukawa matrices of the two different Higgs doublets, so the Yukawa matrices are flavour aligned. Global fits of this scenario in the Bayesian framework~\cite{Karan:2023kyj,Eberhardt:2020dat} (assuming no new complex phases, i.e.\ the alignment parameters are real) show a more general picture of the constraints on the masses than can be gained by studying the specific types. These Bayesian analyses place lower bounds on the Higgs boson masses, e.g.~$M_{H^\pm} \ge 390\gev$ at 95\%~\cite{Karan:2023kyj}. However, these bounds are prior dependent. In particular, assuming Yukawa alignment
or negligible mixing, the mass limits disappear.

Finally, as an alternative to a flavour-conserving 2HDM, one can consider the general 2HDM with no $\mathbb{Z}_2$ symmetry or flavour alignment. This results in general, complex Yukawa couplings and the possibility of tree-level FCNC. Global fits are particularly important in this context, as they can simultaneously test the consistency of many flavour observables with the presence of small, non-zero flavour-violating couplings.
The global fits in \refscite{Athron:2021auq,Athron:2024rir} identified regions that improved the fit to charged-current $B$ meson decays, and $R_{D^{\vphantom{*}}}$ and $R_{D^*}$ anomalies, while remaining consistent with other precision flavour constraints such as $B_s$--$\bar{B}_s$ oscillations. These regions accommodated a sizeable diagonal Yukawa coupling $\rho\mkern-2mu_{\ell}^{\tau\mkern-2mu\tau}$,
and a non-zero flavour-violating Yukawa coupling $\rho_{u}^{t\mkern-1mu c}$ at the $2\sigma$ level.
The specific textures for the Yukawa couplings that are preferred by the global fit provide valuable guidance for model building and have been used to demonstrate that explanations of the $R_{D^{\vphantom{*}}}$ and $R_{D^*}$ anomalies can source sufficient \CP violation to explain the matter-antimatter asymmetry through a successful electroweak baryogenesis mechanism~\cite{Athron:2025iew}. These examples illustrate that a global fit is essential for exploring the interplay between different sectors of the model and for identifying viable, non-standard scenarios that are not apparent when considering individual observables. This is particularly apparent in flavour anomalies, which we now discuss in more detail.

\subsection{Flavour}\label{sec:flavour}

A wealth of new measurements of rare $B$ decays has emerged from flavour factories and LHC experiments in recent years. Some results, especially those involving $b \to s \ell^+\ell^-$ transitions, show intriguing deviations from SM predictions and are often referred to as flavour anomalies. This situation provides a clear illustration of why global fits are essential: with many correlated observables, only a comprehensive analysis can reveal the underlying physics picture.

The most persistent tensions arise in angular observables and branching ratios of exclusive $b \to s \ell^+\ell^-$ transitions, such as $B \to K^{(*)} \ell^+\ell^-$, $B_s \to \phi \ell^+\ell^-$, and $\Lambda_b \to \Lambda \ell^+\ell^-$, as measured by LHCb~\cite{LHCb:2013ghj,LHCb:2014cxe,LHCb:2015tgy,LHCb:2015wdu,LHCb:2015svh,LHCb:2020lmf,LHCb:2020gog,LHCb:2021xxq,LHCb:2021zwz,LHCb:2025update} and CMS~\cite{CMS:2024atz,CMS:2024syx}. Global flavour fits combining the branching ratios and angular observables show large deviations (see e.g.~\refscite{Alguero:2019ptt,Alok:2019ufo,Datta:2019zca,Kowalska:2019ley,Bhom:2020lmk,Biswas:2020uaq,Hurth:2020ehu,Ciuchini:2022wbq,Alguero:2023jeh,Ali:2025xkw} and \refcite{Hurth:2025vfx} for the most recent fit). The CMS measurements of the angular observables in $B \to K^* \mu^+\mu^-$ were found to agree with the corresponding LHCb results to within one standard deviation, providing an independent confirmation of the observed tensions.
The main outcome of the global fits is that the deviations can be consistently accommodated by new-physics contributions to the effective Wilson coefficient $C_9$, corresponding to a modification of the semileptonic vector contribution.
However, SM predictions for these quantities are incomplete, missing non-factorizable power corrections that cannot be calculated yet, which limits the interpretation of these anomalies as clear indications of new physics.
In addition, variations in the form-factor inputs, whether from Light-Cone Sum Rules (LCSR) or lattice calculations, significantly affect the inferred new-physics significance, suggesting that current form-factor uncertainties may still be underestimated~\cite{Chobanova:2017ghn}. Moreover, including the highest low-$q^2$ bins in the global fits produces an artificial enhancement of the new-physics preference, likely reflecting the breakdown of the QCD factorization (QCDf) description in the region just below the charmonium resonances. In a global one-parameter fit to $\delta C_9$, the combined $b \to s\ell\ell$ available data yield a new-physics significance of about $5\sigma$
under the assumption of a 10\% power-correction uncertainty. When inflating this uncertainty to 100\% of the leading non-factorisable amplitude, the significance drops to $\sim2.3\sigma$, underscoring the persistent mismatch between data and QCDf expectations~\cite{Hurth:2025vfx}.

In contrast to the branching ratios and angular observables, tests of lepton flavour universality, including $R_{K^{(*)}}$ and $R_{D^{(*)}}$ measured by the LHCb~\cite{LHCb:2015gmp,LHCb:2017smo,LHCb:2017rln,LHCb:2017avl,LHCb:2019hip,LHCb:2021trn,LHCb:2022qnv}, BaBar~\cite{BaBar:2013mob} and the Belle collaborations~\cite{Belle:2015qfa}, provide theoretically clean probes since hadronic uncertainties largely cancel in these ratios. The latest measurements, particularly for $R_{K^{(*)}}$, are in good agreement with the SM, reducing the overall significance of the anomalies~\cite{LHCb:2022vje}. $R_{D^{(*)}}$ however are currently in tension with the SM prediction at the level of $3.8 \sigma$~\cite{HFLAV2025Moriond}.

Furthermore, recently the branching ratio of $B^+ \to K^+ \nu\bar{\nu}$ has been measured by Belle-II~\cite{Belle-II:2023esi} with a value that is $2.7\sigma$ deviations above the SM prediction.
This growing dataset and evolving experimental landscape underscores the need for coherent global fits capable of consistently treating correlations, theoretical uncertainties, and the interplay among different operator structures in the effective field theory description.

Various proposals were made to explain these anomalies using extensions of the SM such as 2HDMs
(see \cref{sec:thdm}), leptoquark models~\cite{Hiller:2014yaa,Gripaios:2014tna,deMedeirosVarzielas:2015yxm,Sahoo:2015wya,Becirevic:2016oho,Alok:2017jgr,Crivellin:2019dwb,Belanger:2021smw,Aebischer:2022oqe} and $Z'$ models~\cite{Buras:2013dea,Crivellin:2015mga,Darme:2023nsy,Crivellin:2015lwa,Crivellin:2016ejn,Ko:2017yrd,Crivellin:2020oup,Ko:2017quv,Bonilla:2017lsq,Bian:2017rpg,Duan:2018akc,Geng:2018xzd,Ko:2019tts,Allanach:2020kss,Davighi:2021oel,Athron:2023hmz}. Global fits are vital as they check whether corrections to flavour observables spoil predictions for other observables where the SM is consistent with data, and allow one to combine evidence from all the anomalies. In light of this, several tools have been built such as \code{SuperIso}~\cite{Mahmoudi:2007vz,Mahmoudi:2008tp,Mahmoudi:2009zz}, \code{EOS}~\cite{vanDyk:2021sup}, \code{Flavio}~\cite{Straub:2018kue}, and \code{FlavBit}~\cite{GAMBITflavourWorkgroup:2017dbx}. The latter is a \gambit{} module designed to calculate flavour observables using \code{SuperIso-4.1} and likelihoods using \code{HEPLike}~\cite{Bhom:2020bfe}.

\subsection{Dark matter}\label{sec:dm}

The SUSY models in \cref{Sec:SUSY} were motivated on theoretical grounds, though included a solution to the DM problem, with DM explained by the relic abundance of the lightest neutralino. To explore solutions to the DM problem in a more agnostic way, one can instead take a ``bottom-up'' approach by adding one or more new fields to the SM Lagrangian specifically to model a DM particle and its interactions. The simplest example of such a DM model is the scalar singlet model (SSM), which extends the SM with a scalar singlet, $S$, that plays the role of DM. The singlet is stabilized by an ad hoc $\mathbb{Z}_2$-symmetry and interacts with the SM Higgs by a so-called portal interaction, $S^2 H^2$~\cite{Silveira:1985rk,McDonald:1993ex,Burgess:2000yq}. The SSM is very simple since the entire DM phenomenology is dictated by only two parameters: the scalar singlet mass ($m_S$) and the portal coupling ($\lambda_{hS}$). Previous studies have shown that there are distinct viable regions that lead to the correct relic abundance of DM in the SSM~\cite{Cheung:2012xb,GAMBIT:2017gge,Athron:2018ipf}. In global analyses of this model --- which included the relic density measurement, LHC searches for DM, Higgs invisible decays, and direct and indirect detection experiments ---  \refscite{GAMBIT:2017gge,Athron:2018ipf,Balan:2023lwg} found viable regions at $m_S \gtrsim 1\tev$ and $125\gev \lesssim m_S \lesssim 300\gev$, and a narrow resonance region was identified at $m_S \simeq m_h / 2$. The choice of statistical framework affected the viability of the resonance region: the Bayesian fit disfavoured it as it requires the masses to be fine-tuned, whereas frequentist fits do not penalize fine-tuning. Thus, the SSM provides an interesting case study where differences between the Bayesian and frequentist statistical frameworks are important. Depending on the choices of parameters, the SSM could stabilize the electroweak vacuum at high scales and remain perturbative up to the Planck scale.
The non-trivial interplay between low-energy experimental constraints and these desirable UV theoretical properties was investigated using a global fit~\cite{Athron:2018ipf}.
The fit found that only a small region of the SSM parameter space could simultaneously satisfy all known experimental constraints whilst stabilizing the electroweak vacuum and remaining perturbative at high scales~\cite{Athron:2018ipf}.

It is possible to generalize the SSM by changing the spin of the DM candidate that interacts via a Higgs portal, in which case an important role of global fits has been to assess the relative validity of models after imposing all known experimental constraints~\cite{Ellis:2017ndg,GAMBIT:2018eea}.  All Higgs portal models were found to be capable of explaining all of DM and giving a good fit to all data. \Refcite{GAMBIT:2018eea} found that in terms of model selection, the case of vector DM was found to require the most tuning, leading to a slight disfavouring within the Bayesian statistical framework. In the case of fermionic DM, a strong preference was found for including a \CP-violating phase that allows suppression of constraints from direct detection experiments, with odds in favour of \CP violation of the order of 100:1.

A more complete approach to bottom-up DM model building is to generalize the interactions of the hypothetical DM particle using effective field theory (EFT). Dark Matter Effective Field Theories (DMEFTs) provide a model-independent description of possible interactions between the DM and nucleons, quarks or gluons of the SM through high-dimensional contact operators --- notably $d=6$ and $d=7$ operators~\cite{Fan:2010gt,Goodman:2010ku,Fitzpatrick:2012ix,Cheung:2012gi,Catena:2014uqa} --- which are assumed to apply up to some cut-off scale $\Lambda$. Depending on the spin of the DM particle, the number of contact operators between DM and quarks and gluons at $d \leq 7$ can be of the order of $\mathcal{O}(10 \text{--} 50)$~\cite{Brod:2017bsw}, but some of them can have mass or velocity suppressed scattering cross-sections. Global fits of DMEFTs provide comprehensive constraints on these effective contact operators. For instance, \refcite{Liem:2016xpm} performed a global DMEFT fit taking into account all possible $d=6$ gauge-invariant operators between DM and quarks and gluons for the case of real and complex scalar DM candidates and data from the relic density measurement, direct detection experiments, indirect detection experiments, and collider searches. Their main finding is that real scalar DM that interacts only with quarks and gluons cannot be lighter than $100\gev$ due to strong bounds from indirect detection, while complex scalar DM enjoys more flexibility due to the fact that the annihilation mechanism is $p$-wave suppressed.
Another global analysis of a DMEFT with the Dirac DM candidate takes into account possible contact operators at $d=6$ and $d=7$~\cite{GAMBIT:2021rlp}. The results strongly depend on the choice of the cutoff scale $\Lambda$, which was scanned over as a separate parameter, despite the fact that changes in $\Lambda$ can be absorbed into changes in Wilson coefficient, in order to assess the DMEFT validity. If the typical momentum exchange in any given experiment of interest exceeds $\Lambda$ the DMEFT ceases to be a valid physical description of the process, and one would instead need to consider the new field content that generates the operator in a valid UV completion of the theory. The power of a global fit approach in this case is exemplified by the conclusion that, for a DM candidate with mass below $100\gev$, the DMEFT cannot remain valid whilst satisfying all experimental constraints. This means, for example, that the LHC could not see a light DM candidate on its own, but must also be sensitive to the new particle or particles that mediate the interactions between DM and SM particles.

The question of DMEFT validity is particularly relevant in high-energy colliders, where the momentum exchange typically \emph{does} lead to a breakdown in EFT validity. For this reason, so-called \emph{simplified models} were constructed and used as benchmarks for LHC searches for DM~\cite{Abercrombie:2015wmb}. The core idea is to extend the SM with two extra particles: a mediator, produced through the $s$-channel which can be scalar or a vector boson, and a DM particle which can be a fermion, a scalar, or a vector boson. The DM particle interacts with the SM via the mediator; thus, simplified models generalize Higgs portal models using a new mediator as an alternative portal between DM and the SM. Two notable global analyses of simplified $s$-channel DM models with vector mediator have been carried out~\cite{Chang:2022jgo,Chang:2023cki}. Next-generation simplified DM models where the mediator is a coloured particle that is produced through $t$-channel processes are also receiving attention (see e.g.~\refcite{Arina:2023msd} for a recent analysis).

In searches of WIMP-like DM below the GeV scale, collider and direct detection methods become insensitive. This region between 1\,MeV and 1\,GeV, dubbed ``sub-GeV dark matter'' has thus become the focus of increased scrutiny as the window on traditional WIMP candidates closes. This area of parameter space is also attractive, since, as the local dark matter density is known, its flux scales as $1/m_{DM}$. This means that much smaller exposures in direct detection experiments are required, at the cost of a lower required energy threshold. Detector technologies such as skipper CCDs, with excellent sensitivity to electron recoils~\cite{SENSEI:2024yyt,DAMIC:2019dcn}, are among the current leading experiments, but a number of promising quantum technologies have been suggested to tackle this region (e.g.~\refscite{Carney:2022pku,Beckey:2023shi}).    A global analysis of this region~\cite{Balan:2024cmq} focused on a simplified model involving a fermionic DM candidate and a hidden photon mediator, that is, a spin-1 boson that kinematically mixes with the SM photon (for related works see e.g., \refscite{Cheek:2025nul,Wang:2025tdx,Gori:2025jzu}). In this case, couplings and masses required to reproduce the observed relic density lead to strong constraints from electron-recoil direct detection experiments, dark age energy injection altering the CMB, and fixed target experiments. Indeed, NA64 and BaBar have placed strong constraints on invisibly-decaying final states. Taken together, the global analysis yields the startling conclusion that such models can be largely excluded unless a strong asymmetry between dark matter $\chi$ and its antiparticle $\bar \chi$ is present, or if freeze-out is obtained in a resonant region of parameter space. A secondary --- but important --- finding of this work was that benchmark points that have been used as discovery targets for light DM searches at accelerators are actually in tension with the full set of constraints. A new set of benchmark points was thus established as promising targets for future searches.

As the Earth orbits the Sun, any direct signal of dark matter should be accompanied by an \textit{annual modulation}, as the ``wind'' of dark matter is periodically boosted and slowed with respect to the lab. There has been a long-standing claim by DAMA/LIBRA~\cite{Bernabei:2022xgg} of such an annual modulation signal in NaI scintillator consistent with the expectation from particle DM, despite apparent exclusion by total rate experiments using different materials by over six orders of magnitude~\cite{LZ:2024zvo}. Previous global analyses struggled to reconcile these experiments~\cite{Savage:2008er,Kopp:2009qt,Kang:2018qvz,Kang:2019uuj,Tomar:2019urz}. Nonetheless, an experimental program has been developed to test the claim using an identical detector, namely ultrapure sodium iodine. Two recent experiments, COSINE-100~\cite{Carlin:2024maf} and ANAIS-112~\cite{Amare:2025dfq} recently released datasets using NaI, each showing about $4\sigma$ tension with DAMA.
The recent \refcite{Busoni:2025tqa} reexamined the modulation claim in light of a dark matter hypothesis. Specifically, this work compared the expected modulation amplitude seen in each of these three experiments (DAMA, COSINE and ANAIS), given a common nuclear recoil signature. Because of quenching factor and efficiency differences, a signal could indeed be visible at only a subset of experiments. However, the combination of data paint a stark picture: allowing for 6 different low-energy EFT models with 11 free parameters leads to a $5.1\sigma$ tension between DAMA and the combination of ANAIS and COSINE. A second set of analyses using a general, agnostic parametrization of nuclear recoil leads to equally damning results. Bringing the experiments into rough agreement (i.e.\ less than $3 \sigma$ tension) requires significant gymnastics, with bin-by-bin cancellations between recoils on sodium and iodine and negative modulation amplitudes that risk being larger in magnitude than the overall rate.

Finally, the QCD axion or an axion-like particle (ALP) could play the role of DM~\cite{Preskill:1982cy,Abbott:1982af,Dine:1982ah,Turner:1985si}. The QCD axion could solve the strong-\CP problem and more generally ALPs are well-motivated as they are a natural consequence of superstring theories~\cite{Witten:1984dg,Svrcek:2006yi,Arvanitaki:2009fg,Cicoli:2012sz}. Furthermore, as well as playing the role of DM, ALPs could reduce tension in measurements of white dwarf cooling and the Universe's transparency to gamma rays~\cite{Meyer:2013pny,Giannotti:2017hny}. Global fits of axions and ALPs have been performed in Refs.~\cite{Hoof:2018ieb,Athron:2020maw,Balazs:2022tjl} taking into account, amongst other things, data from light-shining-through-wall experiments, helioscopes, haloscopes, the DM relic density measurement, distortions of gamma-ray spectra, Supernova 1987A, Horizontal Branch stars and white-dwarf cooling hints, XENON-1T. For instance, it was found that the QCD axion makes a non-negligible contribution to the DM density in the Universe although it is subdominant~\cite{Hoof:2018ieb} while constraints on the axion mass depend on the underlying model assumptions. Without loss of generality, the axion mass could be anywhere between about $0.1\,\mu$eV and 1\,meV.

\subsection{Neutrino physics}\label{sec:neutrinos}

While in the minimal SM individual lepton flavour numbers are strictly conserved and neutrinos are massless, it has been firmly established by independent solar, atmospheric, reactor and accelerator experiments that neutrinos undergo flavour oscillations. Individual experiments can normally observe only a subset of the possible appearance/disappearance processes and/or a restricted range of neutrino energies and propagation distances; hence,  global fits of neutrino data have been essential in demonstrating that the experimental results can be coherently explained in terms of neutrino quantum mechanical oscillations, driven by a single unitary three-flavour mixing matrix and two independent mass (squared) differences, while the impact of more exotic physics is strongly constrained~\cite{Gonzalez-Garcia:2007dlo}.
Efforts to combine all available neutrino data in a unified statistical analysis date back to the end of 1970s~\cite{DeRujula:1979brg} when the only positive indication for neutrino oscillations was the solar neutrino problem~\cite{Bahcall:1978fa}, even though strong conclusions at that time were hindered by the limited data availability. As more data was collected and also atmospheric experiments pointed to a deficit in the expected ratio of $\nu_\mu$ to $\nu_e$ induced events (atmospheric neutrino anomaly)~\cite{Beier:1992sf}, by the early 1990s it was possible to start constraining some of the oscillation parameters in well defined regions~\cite{Fogli:1993ck}, by combining solar, atmospheric, accelerator and reactor neutrino experiments within a hierarchical three generation scheme~\cite{Fogli:1993ck}. During the 2000s a coherent picture for the values of the neutrino oscillation parameters emerged~\cite{Gonzalez-Garcia:2000opv,Maltoni:2004ei,Fogli:2005cq,Schwetz:2008er}, and notably global fits of neutrino data were able to provide hints~\cite{Fogli:2008jx,Gonzalez-Garcia:2010zke} and evidence~\cite{Fogli:2011qn,Schwetz:2011zk} for the mixing angle $\theta_{13} \neq 0$ before this was experimentally discovered~\cite{DayaBay:2012fng}.
In the following decade, 2010s, global fits were able to start constraining the value of the Dirac \CP-violating phase $\delta$~\cite{Fogli:2012ua,Gonzalez-Garcia:2014bfa} and provide hints for a normal ordering of neutrino masses~\cite{deSalas:2017kay,Esteban:2018azc}.
Currently, neutrino global fits have entered the precision era, with many of the oscillation parameters determined at the percent level~\cite{deSalas:2020pgw,Esteban:2024eli,Capozzi:2025wyn}.
The results of global fits of neutrino data are furthermore used as input in the design of future experiments, in order to maximize their discovery power~\cite{JUNO:2015sjr,Hyper-Kamiokande:2018ofw,DUNE:2020lwj}.

In addition to measurements of neutrino mass squared differences by oscillation experiments, the absolute neutrino mass scale can be inferred from cosmological observations. The neutrino mass determines when neutrinos become non-relativistic, which affects the expansion history of the Universe, the growth of cosmic structure and the cosmic microwave background~(CMB).  While global fits have been present in cosmological work since the early 2000s~\cite{Lewis:2002ah}, these have generally not been combined with likelihoods from earth-based particle physics experiments.  \gambit{} has recently incorporated \code{CosmoBit}~\cite{GAMBITCosmologyWorkgroup:2020htv}, a module that can compute cosmological observables and likelihoods in BSM scenarios. These include big-bang nucleosynthesis, CMB and large scale structure, and type Ia supernova observables. Interfaces to codes like \code{MultiModeCode}~\cite{Price:2014xpa} allow for full inflationary model predictions to be included in \gambit{} scans that encompass a wide range of early Universe and late-time probes. As a flagship example of \code{CosmoBit}'s power, \refcite{GAMBITCosmologyWorkgroup:2020rmf} obtained competitive bounds on the mass of the lightest neutrino.

Many fits of absolute neutrino masses have been performed using cosmological observations, but often under the assumption of mass degeneracy and no mixing~\cite{Wong:2011ip,Lesgourgues:2014zoa,Vagnozzi:2017ovm,Planck:2018vyg,Ivanov:2019hqk}. However, more accurate assessments of the neutrino properties involve fits from neutrino terrestrial experiments, i.e.~oscillation experiments, combined with cosmological data, such as the primordial abundance of light elements, observations from supernovae, baryon acoustic oscillations and data from the CMB power spectrum, in order to derive simultaneous constraints on the absolute neutrino mass scale and the neutrino oscillation data~\cite{Fogli:2004as,Fogli:2008ig,Capozzi:2017ipn,Loureiro:2018pdz,GAMBITCosmologyWorkgroup:2020rmf}. A global fit~\cite{GAMBITCosmologyWorkgroup:2020rmf} with 32 free parameters led to a robust bound on the lightest neutrino mass of 0.037\,eV for the normal ordering, and 0.042\,eV for inverted ordering.

Other than in the minimal three-flavour unitary mixing scheme, global fits in neutrino physics have been employed to explore and derive constraints on further extensions of the SM, such as for instance the presence of additional sterile neutrino light states~\cite{Kopp:2013vaa,Gariazzo:2017fdh,Dentler:2018sju}, deviations from unitarity of the lepton PMNS mixing matrix~\cite{Antusch:2006vwa,Parke:2015goa,Blennow:2016jkn}, constraints on the heavy-neutral leptons appearing in the Type-I seesaw mechanism~\cite{Fernandez-Martinez:2015hxa,deGouvea:2015euy,Fernandez-Martinez:2016lgt,Chrzaszcz:2019inj}, constraints on non-standard neutrino interactions (NSI)~\cite{Biggio:2009nt,Coloma:2019mbs} and bounds on modified neutrino couplings~\cite{Coutinho:2019aiy}.
Other interesting applications have been the determination of the matter potential affecting neutrino propagation~\cite{Gonzalez-Garcia:2013usa} or of the solar model parameters~\cite{Song:2017kvf,Gonzalez-Garcia:2023kva}.

\subsection{Effective field theories}\label{sec:eft}

The DMEFT described in \cref{sec:dm} for DM is an example of an Effective Field Theory (EFT).
EFTs can be used to model BSM physics in a model-independent manner, as they encode BSM physics in Wilson coefficients of non-renormalizable operators. The operators in an EFT are dictated only by SM gauge invariance~\cite{Buchmuller:1985jz} and not by renormalizability; consequently, EFTs are only valid below a cut-off scale $\Lambda$ at which new physics must appear. Null results from direct searches for BSM physics at the LHC suggest a hierarchy between the new physics scale and the electroweak scale. This motivates EFT approaches because an EFT describing such heavy new physics could be valid at the electroweak scale.
The approach is model-independent --- similar to the EWPTs in \cref{sec:ewpts} --- though goes beyond electroweak observables and parametrizes more general new physics effects.

The Standard Model Effective Field Theory (SMEFT) extends the SM by all gauge-invariant operators of dimension $n$, $\mathcal{O}^{(n)}$, starting from dimension five up to dimension $d$ (see e.g.~\refcite{Brivio:2017vri} for a review). The generic form of the Lagrangian becomes
\begin{equation}
	\mathcal{L} = \mathcal{L}_{\rm SM} + \sum_{n=5}^d \sum_i \frac{c^{(n)}_i {\mathcal{O}}_i^{(n)}}{\Lambda^{n-4}}.
\end{equation}
New physics contributions are encoded into the Wilson coefficients, $c$, and the new physics scale, $\Lambda$.
The number of operators depends on assumptions about e.g.~flavour violation. At $d=6$ for the case of three fermion generations there are $2499$ operators that conserve baryon and lepton numbers (see e.g.~\refcite{Alonso:2013hga}).
Thus, the number of Wilson coefficients is a limiting factor for global fits. To reduce the number of Wilson coefficients, one can only consider operators that arise from integrating out specific BSM physics --- e.g.~GUTs, SUSY, models with exotic fermions, etc.~\cite{deBlas:2017xtg}.
Reduced sets of Wilson coefficients at $d=6$ were constrained in global fits that incorporated electroweak precision observables, Higgs boson production and decay measurements, LHC data for top quark and massive gauge-boson production and flavour observables. In the Warsaw basis~\cite{Grzadkowski:2010es}, leading global fits such as those by the \code{SMEFiT} collaboration and other groups have employed Monte Carlo replica methods, nested sampling, and even ML to efficiently constrain the $d=6$ parameter space under specific flavour assumptions~\cite{Ellis:2020unq,Ethier:2021bye}.
In \refcite{Ethier:2021bye}, the global fit considers 36 Wilson coefficients, with the strongest bound being $\Lambda \approx 10$--$20\tev$ using theoretical predictions at next-to-leading order in QCD.

At $d=7$, there are 20 operators if fermion or permutation indices are ignored~\cite{Lehman:2014jma}. This however jumps to 1542 operators for three fermion generations~\cite{Marinissen:2020jmb}.
Although $d=7$ operators are further suppressed by the cut-off scale, they lead to baryon and lepton-number violation phenomena not present in $d=6$ operators. Thus, $d=7$ operators lead to proton decay, neutrino-less double beta decay ($0\nu\beta\beta$), and Majorana neutrino mass terms. However, their predictions can be tested at low-energy and neutrino experiments such as Hyper-K, DUNE and EXO among others along with LHC probes of exotic multilepton signatures. Studies of dimension-seven operators have been performed in e.g.~\refscite{Lehman:2014jma,Liao:2016hru,Liao:2019tep,Zhang:2023kvw,Fridell:2023rtr} while we are not aware of dedicated global analyses. Some of these studies have shown that the scale of new physics can be constrained to be as high as $\Lambda \approx 200\tev$~\cite{Fridell:2023rtr}. Studies of dimension-eight ($d=8$) operators had more focus on Higgs gauge interactions and anomalous quartic gauge couplings (aQGCs)~\cite{Murphy:2020rsh,Li:2020gnx,Corbett:2023qtg,Ellis:2020ljj}. For $d=8$ operators, unitarity and analyticity constraints provide compelling bounds on the corresponding Wilson coefficients, while current and future collider constraints set even stronger bounds that can go up to $\Lambda \approx 16\tev$~\cite{Ellis:2020ljj}.

\section{Future applications, prospects and challenges}\label{sec:future}

\subsection{The precision frontier}

Although the discovery of the Higgs boson was a remarkable achievement by the LHC experiments, ATLAS and CMS, the null results of new particle searches challenged expectations based on naturalness and left fundamental questions unanswered. In light of these challenges, the precision frontier (see e.g.~\refscite{Jaeckel:2010ni,Artuso:2022ouk}) provides an important alternative to direct production of new particles --- an indirect probe of physics beyond the SM due to
virtual effects from heavy states, entering via loop corrections or higher-dimensional operators. This frontier relies on precise experimental measurements (or constraints) of observables whose SM predictions are known to high precision. When theoretical and experimental uncertainties are both tightly controlled these observables can be sensitive to new physics at energy scales well beyond the direct reach of current colliders. If there are significant deviations from the SM, precision calculations of the observables in new physics theories can then facilitate the identification and testing of new physics explanations.

Precision probes include electroweak precision observables (EWPOs),
electric dipole moments~\cite{Pospelov:2005pr}, magnetic dipole moments~\cite{Athron:2025ets, Aoyama:2019ryr, Eidelman:2007sb}, atomic parity violation (APV)~\cite{Marciano:1990dp,Ginges:2003qt},
flavour sector observables and finally, now that we are in the post Higgs era, precision Higgs measurements~\cite{Dawson:2022zbb}.
The growing library of differential cross-section and event shape measurements being published by LHC experiments and predicted by increasingly sophisticated SM calculations can also be considered to be a new class of ``precision observables.'' While the precision is likely to remain at the per-cent rather than per-mille level,  the power of differential measurements will put them on a par with, or much better than, the more inclusive EWPOs in many cases.
Depending on the specific new physics scenario under consideration, these precision observables may constrain physics beyond the mass reach of direct production at colliders or provide complimentary constraints on light new physics that cannot easily be probed by colliders.  Furthermore it can also be important to combine the precision observables as they are sensitive to different features of new physics.

EWPOs are used in the EWPTs that were discussed in \cref{sec:ewpts}, and include e.g.\ the $W$ boson mass and the effective weak mixing angle, $Z$-pole observables such as the total width and asymmetries, and fermion partial widths.  EWPOs have broad sensitivity to new physics that modifies the electroweak sector of the SM at tree-level and/or through radiative corrections~\cite{Peskin:1990zt,Altarelli:1990zd}. Through the $W$ mass, EWPOs are especially sensitive to new physics that violates the custodial symmetry, providing a crucial test for many extensions of the SM Higgs sector~\cite{Sikivie:1980hm, Sirlin:1980nh}.

Originally inspired by the discovery of weak parity violation~\cite{Lee:1956qn,Wu:1957my}, APV measurements~\cite{Zeldovich:1959pn,Bouchiat:1974zz,Barkov:1978bi,Bucksbaum:1981gs,Conti:1979vwd} probe low momentum transfer neutral-current interactions and are particularly sensitive to modifications of the electron–quark weak charges from new physics~\cite{Marciano:1990dp,Ginges:2003qt}. Theoretical predictions rely on atomic-structure calculations combined with precisely measured atomic energy levels~\cite{Roberts:2014bka}. APV probes four-fermion operators and low-energy corrections to the weak mixing angle, and can place competitive limits on certain $Z'$ models~\cite{Ginges:2003qt,Safronova:2017xyt}. APV measurements in combination with collider extraction of $\sin^2\theta_W$ test the running of electroweak couplings over many orders of magnitude~\cite{Erler:2004in,Safronova:2017xyt}.

Electric dipole moments (EDMs) are uniquely powerful probes of sources of \CP violation~\cite{Engel:2013lsa,Roussy:2022cmp,Abel:2020pzs}.
The operators for EDMs are flavour conserving but must be both \CP and chirality violating~\cite{Pospelov:2005pr}, and the SM predictions for them are very small~\cite{Bernreuther:1990jx,Engel:2013lsa,Ema:2022yra,Shabalin:1978rs,Shabalin:1982sg}. This makes these observables directly sensitive to high-scale phases in many BSM theories~\cite{Chupp:2014gka}. EDMs therefore probe mechanisms that could explain the matter–antimatter asymmetry~\cite{Cline:2006ts,Davidson:2008bu,Morrissey:2012db,Chupp:2017rkp} and place strong constraints on new electroweak or scalar sectors~\cite{Olive:2005ru,Chupp:2014gka}. Because a single underlying \CP-violating operator can induce EDMs for leptons, nucleons, nuclei, and diamagnetic atoms with very different relative sizes, a global analysis of multiple systems is typically required to disentangle the operator structure~\cite{Chupp:2014gka, Ramsey-Musolf:2020ndm,Degenkolb:2024eve}.

The magnetic dipole moments (MDMs) of leptons~\cite{Dirac:1928hu,Schwinger:1948iu}, $a_\ell$, are excellent tests of new physics because they have both very precise experimental measurements~\cite{Muong-2:2025xyk,Fan:2022eto} and SM theory predictions~\cite{Aoyama:2020ynm,Aliberti:2025beg}.\footnote{Although the experimental precision of the $a_\tau$ measurement is significantly lower than those of $a_e$ and $a_\mu$~\cite{ATLAS:2022ryk,CMS:2024qjo}, it is expected to significantly increase~\cite{Crivellin:2021spu}.}
In particular the muon magnetic moment~\cite{Stockinger:2006zn,Jegerlehner:2009ry,Keshavarzi:2022kpc} is generally the most sensitive because the muon has the right balance between having a large mass to increase the quantum corrections and a long enough lifetime to be measured with sufficient precision. While MDMs are also generated from new physics that flips the chirality, in contrast to EDMs they are \CP conserving as well as flavour conserving, making them sensitive to a far wider set of new physics scenarios that cannot be simply avoided by not introducing new phases~\cite{Czarnecki:2001pv,Lindner:2016bgg,Athron:2021iuf,Athron:2025ets}.

Flavour observables can also be categorized as precision observables, requiring precise theoretical predictions, often at the NLO order, as well as careful measurements.  They provide  sensitivity to flavour violation (FV) as well as flavour universality violation (FUV), and test mechanisms that could explain the fermion mass hierarchy~\cite{Buras:2005xt}. Many flavour transitions in the SM are loop- and CKM-suppressed via the GIM mechanism~\cite{Glashow:1970gm}, so even very small FV or FUV from new physics can produce detectable deviations from the SM.  See \cref{sec:flavour} for a brief review of the current situation.

Since the discovery of the Higgs boson, Higgs physics has rapidly transitioned to precision measurements.  Precision Higgs observables include the Higgs self-couplings and couplings with fermions and gauge bosons, and provide stringent tests of the scalar sector of the SM. Precision Higgs measurements are an essential target for future colliders many of which can be seen as Higgs factories. For example, the increased precision of the high-luminosity LHC (HL-LHC~\cite{hllhc:2020}) and the clean experimental environment at future lepton colliders such as the International Linear Collider (ILC~\cite{ILC:2013jhg}), Circular Electron-Positron Collider (CEPC~\cite{Ai:2025cpj}), Compact Linear Collider (CLIC~\cite{CLICPhysicsWorkingGroup:2004qvu}), Future Circular Colliders (FCC-ee~\cite{FCC:2018evy} and FCC-hh~\cite{FCC:2018vvp}), and even muon colliders~\cite{Accettura:2023ked} are expected to achieve percent-level precision on the Higgs couplings.

Collider constraints and low-energy experiments can be complementary. For example, understanding sources of \CP violation would provide hints about the mechanisms of baryon asymmetry in the Universe and strongly constrain the possible solutions to this problem~\cite{Barman:2022pip,Gritsan:2022php}.
A well-motivated scenario, \CP violation in the Higgs boson couplings to SM particles~\cite{Alarcon:2022ero} could be constrained at both future colliders and through low-energy experiments searching for permanent electric dipole moments.  These future colliders could also constrain EWPOs. However, taking advantage of these measurements would require a reduction in theoretical uncertainty in EWPO calculations and new MC software~\cite{Blondel:2019qlh}.

Thus the precision frontier is probed by a very large and diverse set of observables, but each one alone gives only a limited view of new physics.
Combining hints from such a range of sources into a coherent picture of new physics demands a global fit. However, learning about new physics at the precision frontier requires precise theoretical calculations and a careful treatment of their uncertainties, beyond that available in current software.  Thus progress in the precision frontier depends on both theoretical and experimental developments, and to maximize the impact of such developments in global fits we will have to carefully understand the uncertainties on these and include them in the fits.

\subsection{Switching from top-down to bottom-up}

Null results from searches for new physics at the LHC and elsewhere have also prompted a major shift in interest from top-down model building to bottom-up approaches. The bottom-up approaches make predictions for weak-scale phenomenology but remain agnostic about physics in the UV. This can be achieved by building a simplified model that includes only the new particles and interactions for a particular collider signature~\cite{Alwall:2008ag,LHCNewPhysicsWorkingGroup:2011mji} or DM annihilation mechanism~\cite{Abdallah:2014hon}, or by modelling unknown interactions without introducing any new states in an effective field theory~(EFT). The interactions in an EFT are \emph{solely} restricted by symmetries and not by renormalizability; the EFT based on the SM field content and SM symmetries is known as the SMEFT \see{Brivio:2017vri,Falkowski:2023hsg,Isidori:2023pyp}.

Simplified models face theoretical challenges, such as gauge invariance and unitarity~\cite{Kahlhoefer:2015bea}. From a phenomenological  perspective, by focusing on such specific signatures, simplified models might miss the opportunity to combine constraints or indeed evidence from multiple searches and experiments. Global fits play an essential role in mitigating this issue using data from multiple experiments in estimates of the current bounds and future discovery prospects. This is witnessed, for example, in recent global fits of simplified DM models with vector mediators. Robust conclusions were not possible unless all possible constraints were properly included in the fit~\cite{Chang:2022jgo,Chang:2023cki}.

The merits of EFT versus simplified models were explored in \refcite{DeSimone:2016fbz}. Compared to simplified models, in the context of global fits EFTs are more complicated due to the complexity of their parameter spaces. For example, in the SMEFT, the number of Wilson coefficients  at $d=6$  can be of the order $\mathcal{O}(3000)$~\cite{Henning:2015alf}. The precise number depends on assumptions about e.g., \CP violation. As a consequence, the curse of dimensionality is a limiting factor for studying the SMEFT in a global fit and restrictive assumptions  must be made.   For example, \refcite{Ellis:2018gqa} fitted 20 $d=6$ operators to over 100 measurements from the LEP, the Tevatron and the LHC. In a similar analysis, \refcite{Ethier:2021bye} forecasted the reach of future colliders in a 19 parameter fit.

Lastly, as discussed in \cref{sec:physics}, the gulf between the new physics scale and electroweak scale leads to computational challenges. EFTs emerge as a potential solution here~\cite{Cohen:2022tir} through  matching and running to UV theories. With the existence of automated tools, such as \code{MatchMakerEFT}~\cite{Carmona:2021xtq} and \code{MATCHETE}~\cite{Fuentes-Martin:2022jrf}, that can provide Wilson coefficients from the matching of new physics models to the EFT up to the one-loop order, global fits will be pivotal tools to pave the way in establishing clear road maps for future studies.

\subsection{Testing as well as forecasting}

Early global fits of BSM physics were primarily focused on forecasting the discovery prospects of new physics at the LHC before its operation, leveraging a combination of data from different experiments. Because of the absence of unambiguous new physics signals at the LHC and elsewhere, interest has somewhat shifted towards testing the goodness-of-fit of models, accumulating evidence from smaller hints of new physics, and cross-checking whether explanations for anomalies are consistent with existing data. For example, after the Higgs discovery, \refscite{Balazs:2012vjy,Bechtle:2015nua} tested the viability of minimal supersymmetric models in Bayesian and frequentist frameworks, respectively. The latter involved computing a $p$-value for the CMSSM. This move to testing the viability of models presents a dramatic new computational challenge on top of already challenging parameter scans, as standard asymptotic results~\cite{Cowan:2010js} are not applicable, and already computationally expensive fits have to be repeated on the results of many toy experiments --- so-called parametric bootstrap in statistics terminology. Another such goodness-of-fit attempt was carried out for the EWMSSM in~\refcite{GAMBIT:2018gjo} based on an approximate approach with less statistical power to detect signals. The construction of goodness-of-fit tests for high-dimensional models with expensive MC likelihoods that show good statistical power remains an important open problem in global fits.

To accumulate evidence from several experiments, global fits could find subtle correlations among a model's parameters and connect hints found in several individual analyses, thus uncovering new excesses with respect to the SM. For example, in a global fit in \refcite{GAMBIT:2018gjo}, a combination of all neutralino and chargino searches at the LHC revealed a $3.3\sigma$ excess when using $13\tev$ data. This happened because small hints could be simultaneously explained by particular combinations of masses and mixings. Secondly, the axion global fit in \refcite{Athron:2020maw} showed that an explanation for excess electron recoil events in a direct search for DM was in tension with astrophysical data. The tension was quantified using both Bayesian and frequentist methods. In this case, the former was computationally challenging as the analysis included several cross-checks on prior dependence. In these cross-checks, fits were repeated with different choices of hyperparameters and priors. Lastly, as discussed in \cref{sec:flavour}, there are several outstanding experimental anomalies in $B$-physics, pointing towards new physics in the flavour sector; global fits allow one to accumulate evidence from these anomalies and cross-check explanations.

Finally, interest in forecasting using global fits could come back around. There are ongoing proposals to build a next-generation collider --- such as an $e^+e^-$, $pp$, and even $\mu^+\mu^-$ collider --- at centre-of-mass energies from the $Z$-boson pole all the way up to $100\tev$. Precision machines such as Higgs and $Z$ factories will provide indirect constraints on the energy scale of new physics up to $\mathcal{O}(10)\tev$ or more, while hadron and muon colliders with center-of-mass energies of up to $100\tev$ can serve as discovery machines. There are several projections of the new-physics reach of these future colliders~\cite{Moortgat-Pick:2015lbx,FCC:2018vvp,FCC:2018evy,AlAli:2021let}. Global fits could play an important role in forecasting the discovery prospects at these future collider experiments --- that is, can they reach the parameter space allowed, or even favoured, by existing experiments? --- see e.g.~\refscite{Athron:2022uzz,Dai:2023pli} for examples.

\subsection{Impact of revolution in Machine Learning}

\begin{figure}
	\centering
	\includegraphics[width=0.95\linewidth]{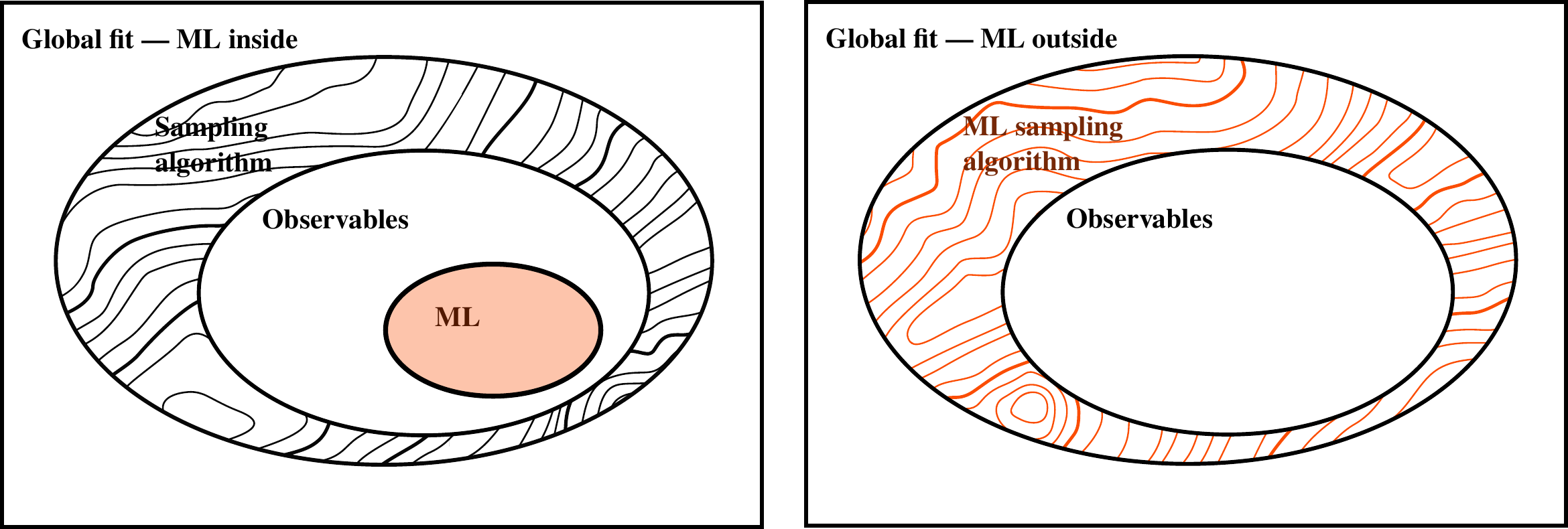}
	\caption{ML could play a role in global fits by speeding up the computation of expensive observables inside a fit using a traditional sampling algorithm (left) e.g.~event generation, or by driving the entire fit, replacing traditional sampling algorithms (right).}
	\label{fig:role}
\end{figure}

Because of the computational cost of global fits, ML methods were applied in global fits as early as 2010~\cite{Bridges:2010de}. Fits were repeated numerous times to explore the coverage properties of confidence intervals. Repetition was accelerated by learning the mapping between fundamental parameters and the particle mass spectrum, leading to a speedup by a factor of about $10^4$. These techniques were packaged into the general purpose \code{BAMBI} code~\cite{Graff:2011gv}, which combined neural networks and the nested sampling algorithm. The neural network learned a computationally cheap emulator for an expensive likelihood call. Since then, ML methods have been pioneered for both statistical computation and physical modelling of observables. Consider, for example, normalizing flows~\seeone{Papamakarios:2019fms}, which transform simple probability distributions into complex ones using a sequence of invertible transformations, enabling efficient density estimation and sampling. These flows could be used in global fits in several ways: they were proposed in \refcite{Williams:2021qyt} as a subalgorithm inside nested sampling for Bayesian computation; in
\refscite{Gao:2020vdv,Albergo:2021bna} as the basis of novel algorithms for statistical computation; and in \refcite{Krause:2021ilc} for modelling calorimeter showers. Thus, the role of ML inside global fits remains an open question, as illustrated in \cref{fig:role}.

A hallmark of ML methods is a focus on generation of new data from training data, e.g., generation of text after training on a text corpus, such as \code{ChatGPT}~\cite{OpenAI:2023ktj} --- these are generative models. In physics, generative models can create realistic synthetic data by modelling the underlying probability distribution of experimental data or detailed computer simulations. There are three common types of generative models:
\begin{enumerate*}
	\item generative adversarial networks (GANs), where two neural networks compete against each other. The first network attempts to generate realistic new samples and the other one must discriminate them from the original samples.
	\item variational autoencoders (VAEs) that perform a lossy encoding of the data into latent space and then decode that to generate variations.
	\item diffusion models, which transform noise into structured data through iterative refinement of the initial solution.
\end{enumerate*}
GANs~\cite{Rodriguez:2018mjb,Paganini:2017dwg,Butter:2020qhk,Krachmalnicoff:2020rln,Buhmann:2023pmh}, VAEs~\cite{Otten:2019hhl} and normalizing flows~\cite{Gao:2020zvv} have all already been applied to the problem of Monte Carlo (MC) event generation (see also \refscite{Butter:2022rso,Alanazi:2021grv} for reviews). Thus, ML could be used inside global fits for event generation, removing a major bottleneck. See \refcite{Matchev:2020tbw}, however, for scepticism about data amplification in event generation.

On the other hand, ML could be used to drive a global fit using
a hybrid method that combines ML and a traditional algorithm or a completely new ML algorithm e.g.~\refscite{Caron:2016hib,Ren:2017ymm} which were developed especially for BSM physics. A common recipe for constructing a hybrid method is to keep a standard sampler at the core (e.g.~MCMC or nested sampling) but use ML --- e.g.~flows and flow matching~\cite{Albergo:2019eim,Kanwar:2020xzo,Boyda:2020hsi,Williams:2021qyt,wong2022} or diffusion~\cite{Hunt-Smith:2023ccp} --- to make efficient proposals, perhaps including occasional long jumps. These long jumps could decrease the autocorrelation length of the chain and thus improve sampling efficiency~\cite{Albert:2024zsh}.
ML methods are particularly powerful in simulation-based inference (SBI), which can be used even in cases where the likelihood function is intractable~\seeone{Cranmer:2019eaq}. The earliest form of SBI was Approximate Bayesian Computation (ABC;~\cite{Diggle1984}). In the most basic form of ABC, parameters and data are simulated from the model and accepted if the simulated data matches the observed data to within a user-defined tolerance. Although ABC methods have been used widely in cosmology and astrophysics~\cite{Akeret:2015uha,Ishida:2015wla,Witzel:2018kzq,Baxter:2021tui}, for high-dimensional datasets (or summary statistics) they suffer from the curse of dimensionality, as in high dimensions simulated data almost never matches the observed data. To somewhat alleviate the curse of dimensionality, neural conditional density approximation approaches that use ML have emerged, such as neural posterior estimation (NPE;~\cite{Papamakarios:2016ctj,pmlr-v97-greenberg19a}), neural likelihood estimation (NLE;~\cite{pmlr-v89-papamakarios19a}) and neural ratio estimation (NRE;~\cite{Cranmer:2015bka,thomas2022l}). These techniques use normalizing flows or other neural network architecture to learn the posterior or likelihood from forward simulations and have been used in cosmology, astrophysics and EFTs~\cite{Dax:2021myb,Leyde:2023iof,Brehmer:2018kdj}. Because these are one-shot methods, the computational cost may be amortized as a trained model can be reused for alternative datasets.
Furthermore, problems in the calculation can be diagnosed through simulation based calibration~\cite{talts2020}. The promising statistical properties and the way that the curse of dimensionality manifests in these methods are beginning to be understood~\cite{frazier2024statisticalaccuracyneuralposterior}.

The revolution in ML is connected to hardware developments, particularly single-instruction multiple data (SIMD) on Graphics Processing Units (GPUs), and software that can take advantage of it, e.g., \code{CUDA}~\cite{cuda}, \code{JAX}~\cite{jax}, \code{TensorFlow}~\cite{tensorflow} and \code{Torch}~\cite{torch}. The software developments include
differentiable programming. Despite progress~\cite{Heinrich:2022xfa,MODE:2022znx}, especially in cosmology~\cite{Campagne:2023ter,Piras:2024dml,Balkenhol:2024sbv}, global fits are yet to take full advantage of automatic differentiation. This would enable faster optimization algorithms using gradients as well as more efficient sampling algorithms such as Hamiltonian Monte Carlo~(HMC;~\cite{Neal:2011mrf,Betancourt:2017ebh}). A step in this direction was recently taken in \refcite{AbdusSalam:2025the}. Approximate and differentiable analytic expressions for observables in the CMSSM were constructed using symbolic regression~\cite{Makke:2022rnq} and used in a global fit using HMC. GPUs have the potential to dramatically speedup likelihood evaluations and MC simulations of multi-parton processes at high-energy colliders and thus statistical global fits in the future. There are, however, challenges to taking advantage of GPUs to speed up statistical computation \seeone{Sountsov:2024zwe} or physical modelling.  As discussed, MC simulations usually rely on importance sampling and are computationally expensive, especially at NLO.
GPUs could speed this up~\see{Carrazza:2021gpx,Bothmann:2021nch,Bothmann:2023gew}. In \refcite{Carrazza:2021gpx}, simulation of one million $t\bar{t}gg$ events at LO required about 2500 seconds on the CPU and only about 100 seconds on the GPU.\footnote{Specifically, a \code{AMD 2990WX} CPU with 32 cores and 128 GB of RAM and a \code{NVIDIA RTX 2080 Ti} GPU with 12 GB of VRAM.}
Further improvements are foreseen e.g.~by replacing phase-space samplers with better ones based on process topologies, and speeding-up phase-space integration using ML.

Finally, global fits require software development in multiple programming languages to implement new experimental searches, models and calculations; data wrangling of both experimental data and samples from a global fit; maintenance to keep this software working; and testing and debugging on multiple computer platforms so that teams of researchers can use the software on their computers and high-performance clusters. Large-language models and agentic AI have the potential to rapidly accelerate these time-consuming activities~\see{jalil2023,wang2025,roychoudhury2025,luo2025}. Beyond helping with existing activities, these technologies could automate the conversion of legacy software into languages that support modern GPUs and automatic differentiation. \Refcite{dhruv2024} presents an example of converting the LHC parton-level MC program \code{MCFM}~\cite{Campbell:2010ff} from \code{Fortran} to \code{C++}.

In summary, the potential of the ML has yet to be fully realized in global fits. Although many of the ingredients are available, the relationship between global fits and ML remains unclear.

\section{Summary}\label{sec:summary}

Global fits have played a pivotal role in the exploration of physics beyond the Standard Model (BSM), beginning with early studies of minimal supersymmetric models that anticipated the discovery potential of the LHC. After the LHC began collecting data in 2010, the role of global fits became increasingly crucial in guiding the search for new physics. Incorporating results from dark-matter searches and other experiments, global fits expanded to study not only supersymmetric frameworks but also a broad range of non-supersymmetric theories, including effective field theories (EFTs), dark-matter models, extended Higgs sectors, and models containing axions or axion-like particles. The initial aim of global fits was to determine a model's preferred parameters by intelligently combining constraints from diverse experiments and efficiently scanning its parameter space. More recently, however, their focus has shifted toward model comparison identifying which theoretical framework provides the best description of the data. At the same time, global fits remain indispensable within the Standard Model itself, providing precision tests that may reveal subtle hints of new physics through e.g.~electroweak observables and Higgs-boson rate measurements.
Since no ongoing experiment has yet shown definitive evidence of BSM physics, global fits will continue to play a central role in shaping future directions. They are essential tools on the precision frontier, encompassing electroweak precision measurements, Higgs observables, and a variety of low-energy observables.
The shift from UV-complete models to EFTs further highlights their importance.
Moreover, future hadron and lepton colliders use global fits to evaluate the discovery potential of well-motivated theories.
Finally, the ongoing AI revolution promises to significantly accelerate and enhance global-fit methodologies, through advances such as faster samplers, likelihood-free inference, generative modelling for simulation, improved precision calculations of collider observables, and GPU-accelerated matrix-element calculations.

\section*{Acknowledgements}

We would like to Will Handley, Torsten Bringmann and Felix Kahlhoefer for valuable discussions.

The work of PA and LW is supported by the National Natural Science Foundation of China (NNSFC) Key Projects grant No.\ 12335005.
CB acknowledges support from the Australian Research Council through projects DP220100643 and LE250100010.
AF was supported by the National Natural Science Foundation of China (NNSFC) RFIS-II W2432006.
The work of AJ is supported by the Institute for Basic Science (IBS) under Project Codes IBS-R018-D1 and IBS-R018-D3.
AK, CC and AR were supported by the Research Council of Norway through the FRIPRO grant 323985 PLUMBIN'.
ML is funded by the European Union under the Horizon Europe's Marie Sklodowska-Curie project 101068791 — NuBridge.
R.\ RdA is supported by PID2020-113644GB-I00 from the Spanish Ministerio de Ciencia e Innovaci\'on and by PROMETEO/2022/69 funded by the Generalitat Valenciana.
CS is supported by Project W2441004 supported by the National Natural Science Foundation of China.
WS is supported by the Natural Science Foundation of China (NSFC) under grant numbers 12305115.
MJW is supported by the Australian Research Council grants CE200100008 and DP220100007.

\bibliographystyle{JHEP}
\bibliography{biblio.bib}

\end{document}